\documentclass[12pm, twocolumn]{aastex631}
\usepackage{amsmath}

\newcommand\kms{km$\,$s$^{-1}$\,}
\newcommand\Msol{M$_{\odot}$}

\newcommand{\hi}{H\,{\sc i} }

\newcommand{\halpha}{H$\alpha$ }
\defcitealias{Zibettis}{Z09}
\defcitealias{Taylor+2011}{T11}
\usepackage{booktabs} 

\begin{document}

\title{Citizen Science Identification of Isolated Blue Stellar Systems in the Virgo cluster\footnote{Based on observations obtained with the Hobby-Eberly Telescope (HET), which is a joint project of the University of Texas at Austin, the Pennsylvania State University, Ludwig-Maximillians-Universitaet Muenchen, and Georg-August Universitaet Goettingen. The HET is named in honor of its principal benefactors, William P. Hobby and Robert E. Eberly.}}

\author[0009-0006-0732-3031]{Swapnaneel Dey}
\affiliation{Astronomy Department, Steward Observatory, University of Arizona \\
 933 N Cherry Ave, Tucson, AZ 85719, USA\\}

\author[0000-0002-5434-4904]{Michael G. Jones}
\affiliation{Astronomy Department, Steward Observatory, University of Arizona \\
 933 N Cherry Ave, Tucson, AZ 85719, USA\\}
\author[0000-0003-4102-380X]{David J. Sand}
\affiliation{Astronomy Department, Steward Observatory, University of Arizona \\
 933 N Cherry Ave, Tucson, AZ 85719, USA\\}

\author[0009-0005-9612-4722]{Nicolas Mazziotti}
\affiliation{Astronomy Department, Steward Observatory, University of Arizona \\
 933 N Cherry Ave, Tucson, AZ 85719, USA\\}

\author[0000-0001-9165-8905]{Steven Janowiecki}
\affiliation{University of Texas, Hobby–Eberly Telescope, McDonald Observatory, TX 79734, USA}

\author[0000-0003-2307-0629]{Gregory R. Zeimann}
\affiliation{University of Texas, Hobby–Eberly Telescope, McDonald Observatory, TX 79734, USA}

\author[0000-0001-8354-7279]{Paul Bennet}
\affiliation{Space Telescope Science Institute, 3700 San Martin Drive, Baltimore, MD 21218, USA}

\begin{abstract}
We present a catalog of 34 new candidate (13 high confidence) isolated, young stellar systems within the Virgo galaxy cluster identified through a citizen science search of public optical and ultraviolet imaging. ``Blue blobs" are a class of blue, faint, isolated, extremely low stellar mass, and metal-rich star-forming clouds embedded in the hot intracluster medium of the Virgo cluster. Only six blue blobs were known previously and here we confirm an additional six of our candidates through velocity and metallicity measurements from follow-up optical spectroscopy on the Hobby-Eberly Telescope (HET). Our 13 high confidence candidates (including the six confirmed) have properties consistent with prior known blue blobs and are inconsistent with being low-mass galaxies. Most candidates are concentrated in relatively dense regions, roughly following filamentary structures within the cluster, but avoiding its center. Three of our candidates are likely the stellar counterparts of known `optically dark' clouds of neutral hydrogen in the cluster, while a further four are widely separated extensions to previously known blue blobs. The properties of our new candidates are consistent with previous conclusions that blue blobs likely originated from ram pressure stripping events, however, their locations in velocity--projected cluster-centric radius phase-space imply that their parent galaxies are not on their first infall into the cluster. Through our ongoing follow-up program with HET we aim to confirm additional candidates, however, detailed understanding of the stellar populations and star formation histories of blue blobs will require JWST observations. 
\end{abstract}

\keywords{Star forming regions (1565); Virgo cluster (1772); Low surface brightness galaxies (940); Ram pressure stripped tails (2126); Dwarf galaxies (416)}

\section{Introduction} 
\label{sec:intro}

In galaxy clusters, ram pressure stripping \citep{Gunn+1972} is a ubiquitous process \citep[e.g.][]{Chung+2007,Boselli+2018,Poggianti+2019,Boselli+2022,Roberts+2022} that drives the evolution of infalling star-forming galaxies by pushing out their gas reservoirs and eventually quenching star formation \citep[e.g.][]{Abadi+1999,Vollmer+2001,Solanes+2001,Tonnesen+2007,Crowl+2008,Bahe+2015,Cortese+2021}. In the intermediate stage after ram pressure stripping begins, but before quenching, galaxies often exhibit dramatic ``jellyfish" structures with tendrils of gas and star-formation stretching out in their wake \citep[e.g.][]{Kenney+2004,Ramatsoku+2019,George+2018}. In some cases, dense clumps of intense star formation are visible, sometimes referred to as ``fireballs" \citep[e.g.][]{Yoshida+2008,Jachym+2019}. In others, long (10s of kpc) trails of hot, shocked gas can be identified with H$\alpha$ imaging \citep[e.g.][]{Kenney+1999,Yoshida+2002,Boselli+2018b}. IC3418 in the Virgo cluster presents one of the clearest examples of both traits \citep{Kenney+2014}.
Similarly large features are also commonly identified in cluster galaxies from continuum radio synchrotron emission in tails of hot gas \citep[e.g.][]{Gavazzi+1987,Vollmer+2004,Chen+2020}, and \hi radio spectral line emission in those with cool, neutral gas \citep[e.g.][]{Chung+2007,Ramatsoku+2019}.

Despite the scale of these jellyfish structures (up to $\sim$100~kpc) they are generally connected/associated with a parent galaxy. However, recently \citet{Jones2022} identified a new class of objects, referred to as ``blue blobs," that are typically isolated from any galaxy by 100s of kpc. These objects are actively star-forming, with both UV and H$\alpha$ emission, extremely low stellar mass ($<10^5$~\Msol), made up seemingly entirely of young ($<$200~Myr) stars, metal-rich ($12 + \log \mathrm{O/H} > 8.2$), and have radial velocities \citep[$-500 < cz_\odot/\mathrm{km\,s^{-1}} < 3000$, e.g.][]{Mei+2007} consistent with Virgo cluster membership \citep{Beccari+2017,Sand+2017,Bellazzini+2018,Jones2022a,Jones2022,Bellazzini+2022}. 

Given these properties, blue blobs are likely the ram pressure stripping equivalents of tidal dwarf galaxies (TDGs), that is, stellar systems formed out of pre-enriched stripped gas. \citet{Jones2022} argued that their degree of isolation is only consistent with the young ages of their stellar populations if they are traveling at high speed ($>$1000~\kms) relative to their parent galaxies, especially given that they also reside in the hot intracluster medium (ICM) where they presumably cannot survive long term \citep[though perhaps up to a Gyr;][]{Calura+2020}. Such velocities would be difficult or impossible for a TDG to obtain in a tidal interaction \citep{Bournaud+2006}, and these properties therefore point to ram pressure stripping as the likely formation mechanism. This would make them closely related to fireballs and jellyfish, but representing the disconnected extremities of these structures. In most cases \citet{Jones2022} were unable to confidently identify the parent galaxy because of the isolation of the blue blob in question. 

The first blue blob discovered was identified serendipitously through a search targeting optically ``dark" compact \hi clouds \citep{Adams+2013} thought to be candidate Local Group dwarf galaxies. However, after an optical counterpart of AGC~226067 (also called SECCO~1) was identified \citep{Bellazzini+2015, Sand+2015, Adams+2015} it was quickly realized that this object was far more distant, and likely resides in the Virgo cluster (16.5~Mpc). Follow-up Hubble Space Telescope (HST) imaging supported this conclusion and showed that its visible stellar population consisted entirely of young stars \citep{Sand+2017}, while integral field spectroscopy with the Multi Unit Spectroscopic Explorer (MUSE) on the Very Large Telescope (VLT) showed that the young stars were at the same radial velocity as the \hi \ gas and that it was surprisingly metal-rich \citep{Beccari+2017}. A handful of further examples of blue blobs were subsequently identified through an intensive visual search based on Next Generation Virgo cluster \citep[NGVS,][]{Ferrarese+2012} and Galaxy Evolution Explorer \citep[GALEX,][]{Martin+2005} images \citep{Sand+2017}. Follow-up observations confirmed that these also had properties broadly consistent with SECCO~1 \citep{Jones2022a,Jones2022,Bellazzini+2022,JoneBC6}, indicating that blue blobs are a population of objects, rather than SECCO~1 being a lone enigma.

Identifying and characterizing this population is key to understanding their overall properties and likely formation pathways. Furthermore, they appear to be inconsistent with current simulations of ram pressure stripping in clusters and likely hold important clues for understanding this critical phenomenon \citep[][and references therein]{Jones2022}. Although several of the initial blue blobs were first identified through their \hi \ line emission, \citet{Jones2022} found that not all blue blobs are \hi-rich and thus an optical/UV search is required to create an unbiased sample. Unfortunately, these objects are faint, have low surface brightness, and are highly irregular, making them challenging to identify with traditional algorithms. There are also likely too few known to make machine learning a viable option, as the training sample would be inadequate. This leaves visual classification as the primary means to identify them at present.

In this paper, we present the results of an extensive Zooniverse\footnote{\url{www.zooniverse.org}} citizen science search covering the entire Virgo cluster using archival optical and UV imaging that takes advantage of the brighter appearance of blue blobs in $u$-band and UV. Through this search, we have identified 13 new high confidence blue blob candidates, six of which we have already confirmed with follow-up observations, and 21 lower confidence candidates. In addition, we have identified numerous examples of likely jellyfish structures connected to Virgo member galaxies. 

In the following section, we discuss the search methodology and candidate classification. In \S\ref{sec:HETdata} we describe our optical spectroscopy follow-up program. In \S\ref{sec:starsgas} we outline how we estimate the stellar and gas properties of our candidates and in \S\ref{sec:locations} we discuss their location within the cluster and associations with known objects. In \S\ref{sec:stellarprops} \& \S\ref{sec:rpsloc} we contrast the properties of our candidates with those of low-mass, star-forming galaxies, and discuss their likely points of origin in general. Finally, we present our conclusions in \S\ref{sec:conclusions}.

\section{Citizen Science Search}
\label{sec:method}

In this section we describe the optical and UV images used for the search for blue blobs, outline the search process that volunteers performed, and define how the resulting candidates were classified.

\subsection{Optical \& UV images}

\begin{figure*}[ht!]
\centering
\includegraphics[width=0.7\textwidth]{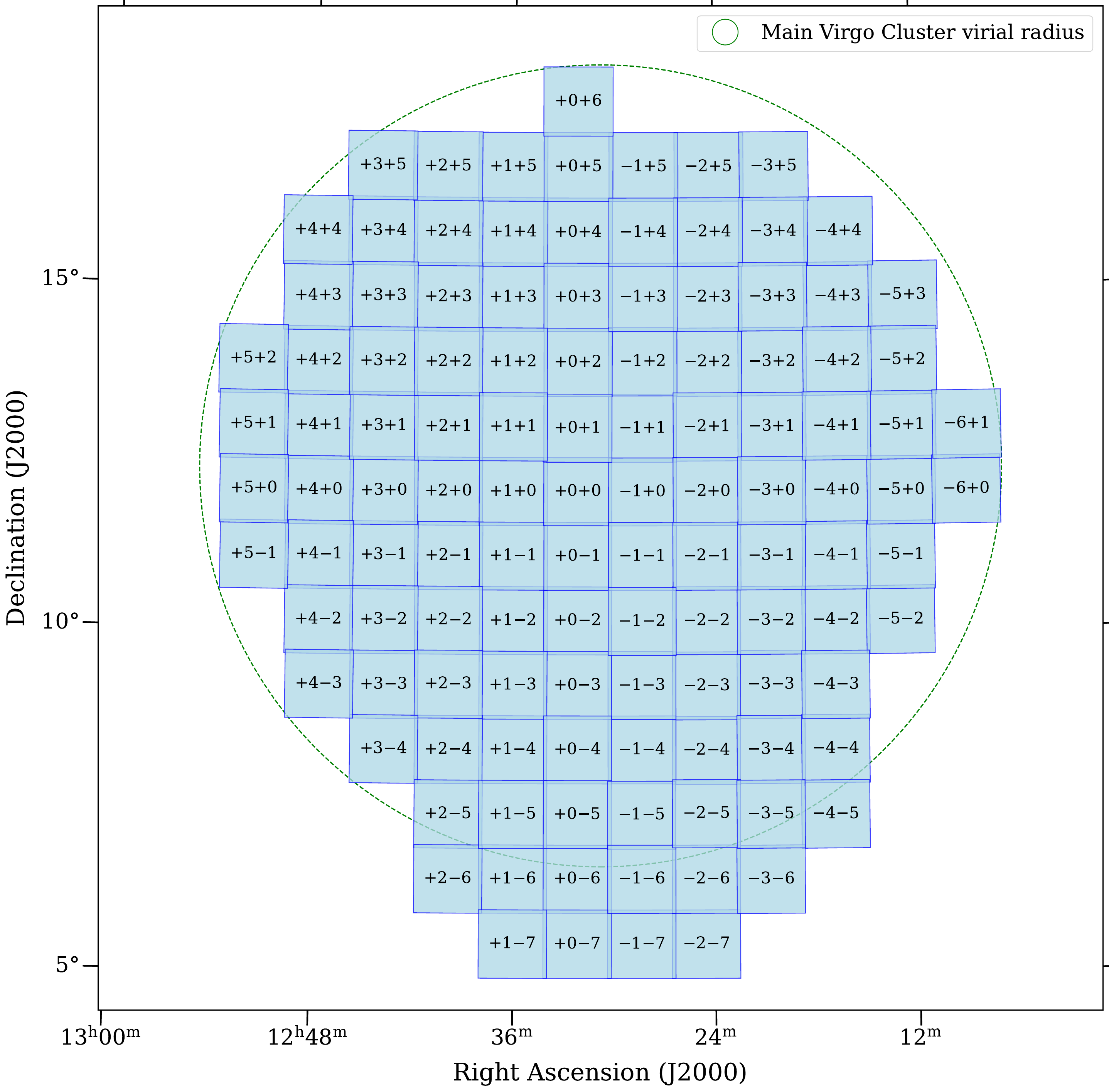}
\caption{NGVS tiles (blue boxes) overlaid on the area enclosed by the main Virgo cluster virial radius (green). The number in each tile is its associated tile number, which indicates its offset (in degrees) from the central tile in RA and Dec. The NGVS coverage extends towards the Virgo B cloud (in the south), but lacks coverage right at the virial radius in most other regions. Each of these tiles was split into 400 overlapping cutouts for the citizen science search.}
\label{fig:tiles}
\end{figure*}

Our search for new ``blue blob'' candidates relies on optical images from the NGVS \citep{Ferrarese+2012}, the Dark Energy Camera Legacy Survey \citep[DECaLS,][]{Dey+2019} and UV images from GALEX \citep{Martin+2005}. NGVS was a deep survey of the entire Virgo cluster spanning approximately 100~sq~deg (see Figure~\ref{fig:tiles}), performed with the MegaCam 1~sq~deg imager on the Canada-France-Hawaii telescope (CFHT). Images were taken in $u^\ast griz$ bands between 2009 and 2013, with a typical point source depth of $m_g = 25.9$ mag. The survey footprint is split up into $\sim$1~sq~deg tiles, which are publicly available from the Canadian Astronomy Data Centre\footnote{\url{www.cadc-ccda.hia-iha.nrc-cnrc.gc.ca}} (CADC). In the case of DECaLS and GALEX, we rely on the $gri$ and FUV+NUV mosaics for the DESI (Dark Energy Spectroscopic Instrument) legacy imaging surveys, which is accessed via the Legacy Viewer\footnote{\url{www.legacysurvey.org/viewer}} cutout service.

We used a \texttt{Python} script to iterate through all the NGVS 1~sq~deg tiles (shown as blue boxes in Figure \ref{fig:tiles}) and split them each into 400 overlapping cutouts, approximately 3\arcmin \ across ($512 \times 512$ pixels with a pixel scale of 0.37\arcsec. The characteristic appearance of blue blobs is that they are faint, very blue, and clumpy. We therefore produced color RGB images of each cutout using the bluest filters ($u^\ast$ and $g$) and the $i$-band (which has more complete coverage over the survey footprint than $r$-band). In particular, using $u^\ast$-band as the blue channel in the RGB image ensures that only the bluest objects appear as blue in the cutouts. In addition to producing the optical RGB cutouts, our script automatically retrieved GALEX NUV+FUV and DECaLS images (with identical fields of view to the NGVS cutouts) from the Legacy Survey viewer. 

\subsection{Search process}

\begin{figure*}
    \centering
    \includegraphics[width=1\linewidth]{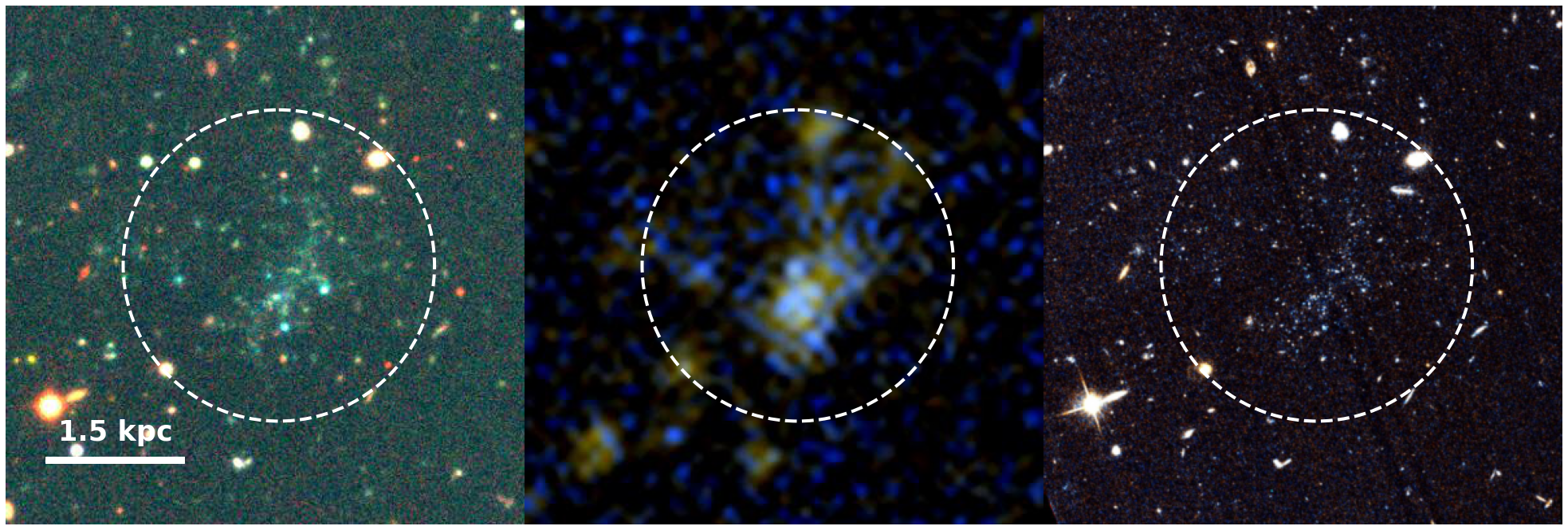}
    \caption{NGVS (left) and GALEX (middle) cutouts and HST imaging (right) of rank 1 candidate, BC1 \citep{Jones2022}. There is a clear presence of blue clumps in the NGVS cutout. It has a strong corresponding UV emission in the GALEX cutout. It is very irregular in shape, and using Legacy Viewer, it was confirmed that this candidate was not associated with any nearby galaxy.}
    \label{fig:BC1 HST}
\end{figure*}

Few blue blobs are currently known, so automated search approaches are poorly suited to their identification. For example, there are too few known objects to train a convolutional neural network. Additionally, blue blobs are distinctive, fairly unique, and varied in shape and would likely pose a challenge for traditional automated algorithms to identify. Blue blobs, like that in Figure~\ref{fig:BC1 HST}, can be faint and hard to spot in optical images. However, they consist of star-forming regions, so they have bright UV emission, making them distinctly identifiable in GALEX images.  For these reasons, visual identification is currently the best-suited search method. 

To streamline the process of identifying candidates by multiple people we used the citizen science platform Zooniverse, which offers a user-friendly interface and automatically tabulates all classifications, making it an efficient tool for projects that involve visual classification of large quantities of data. We began by uploading a subset of the $\sim$150,000 NGVS, DECaLS, and GALEX cutouts, which were then searched by members of our research team. This allowed us to identify additional examples of blue blob candidates, as well as various spurious sources, that could be used as tutorial materials when we opened the project to the public. 

The “Blobs and Blurs: Extreme Galaxies in Clusters” project\footnote{\url{https://www.zooniverse.org/projects/mike-dot-jones-dot-astro/blobs-and-blurs-extreme-galaxies-in-clusters}} was launched to Zooniverse in June 2023 to initially search for blue blobs and diffuse galaxies in the Fornax cluster (Mazziotti et al. (in prep.)). The search did not reveal any blue blobs in the Fornax Cluster. This might be attributed to the Fornax Cluster being older than the Virgo cluster or having a lower mass. Nonetheless, we used the same framework and uploaded all 150,000 cutouts from the Virgo cluster for citizen scientists to examine. Zooniverse has supported many prominent projects \citep[e.g., Gravity Zoo, Active Asteroids, Galaxy Spy;][]{GZ, AS, Galaxyspy} that would be challenging or impossible to complete with any other approach.

\begin{figure}
    \centering  \includegraphics[width=1\linewidth]{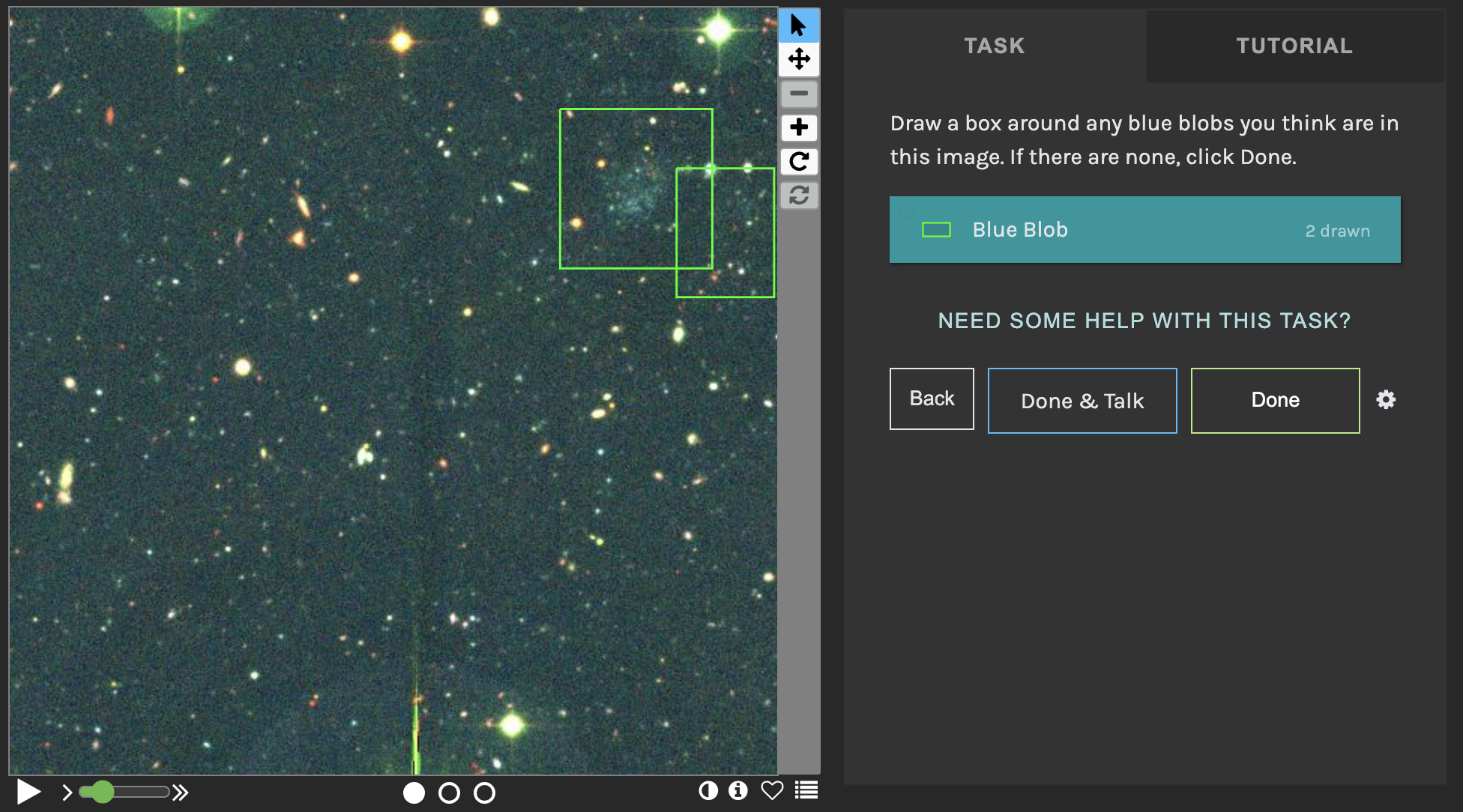}  \includegraphics[width=1\linewidth]{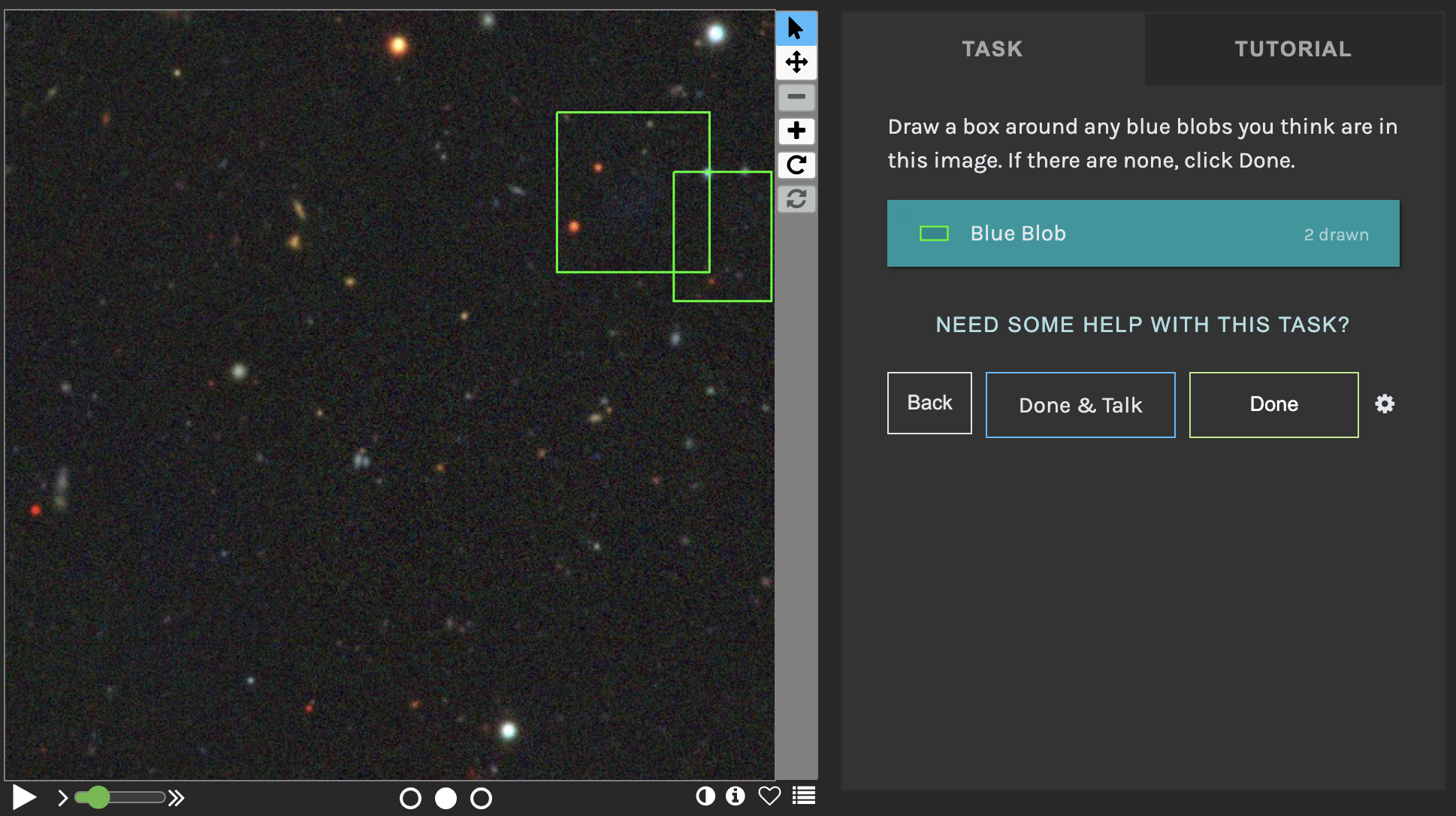} \includegraphics[width=1\linewidth]{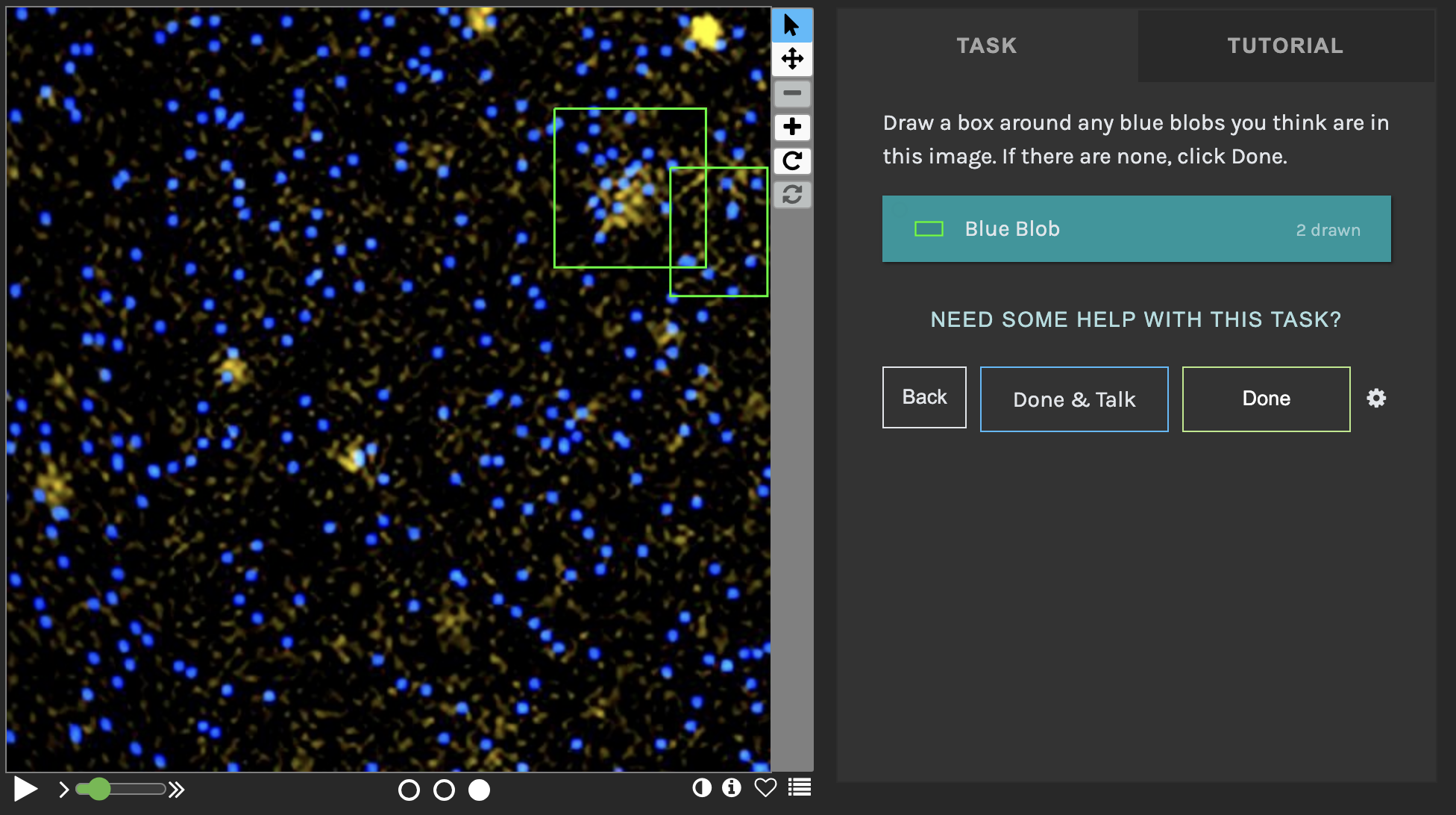}
    \caption{Illustration of the blue blob shown in Figure~\ref{fig:BC1 HST} as displayed on the citizen science platform, Zooniverse. The volunteer flips through NGVS (top), DECaLS (middle), and GALEX (bottom) image cutouts using the dots at the bottom of each cutout. The volunteer can draw multiple green boxes over the blue blob candidate. The volunteer also has the option to discuss the specific cutout with other volunteers further or move on to the next cutout.}
    \label{fig:zooniverse}
\end{figure}

For each cutout, volunteers examined a set of NGVS, DECaLS, and GALEX images, all covering the same field of view within a random part of the Virgo cluster. Note that the cutouts were not centered on any pre-selected candidates, but simply tiled the entire NGVS footprint. Volunteers visually inspected these images to identify potential candidates. They were provided a tutorial and a field guide describing what to look for. The tutorial provided information on how to use the Zooniverse interface to identify and collect information. The first section of the field guide presented images of blue blobs in NGVS, DECaLS, and GALEX and described the appearance of blue blobs in the three different image types. Another section provided images of various blue blobs that the team had discovered before the launch. The field guide also provided information on the common artifacts observed in each type of image. The volunteers were told to identify any object with a blue, clumpy, and irregular appearance accompanied by UV emission in GALEX. They were also advised to consult the field guide frequently to better understand the classification they were attempting.
After 10 volunteers had inspected the same image, it was retired and was no longer shown to additional volunteers. As most cutouts contained only Virgo cluster galaxies, background galaxies, and foreground stars, an image counted as having been inspected when a volunteer viewed it and then clicked ``Done" (Figure~\ref{fig:zooniverse}), even if no object in the image was marked.

To identify a blue blob candidate (BC), volunteers used the Zooniverse draw tool to form one or multiple rectangular boxes around the suspected blue blob in the image (Figure\,\ref{fig:zooniverse}). This classification was then stored in the Zooniverse database for our project. The volunteers could also discuss any cutouts they came across with other people on the project (including our science team). These discussion forums were public and acted as additional training resources for volunteers on the project.

In addition, the searchers were also tasked with marking any diffuse galaxies they encountered.
While blue blobs can be prominent in NGVS and GALEX, they are typically very faint in DECaLS, as seen in the middle cutout of Figure \ref{fig:zooniverse}. We provided volunteers with the DECaLS cutout primarily to aid with the optical identification of diffuse galaxies. However, this study focuses exclusively on blue blobs, with the classification of diffuse galaxies to be covered in a forthcoming publication. Therefore, we will not discuss the DECaLS images further.

We used the \texttt{Panoptes aggregation tool}\footnote{https://aggregation-caesar.zooniverse.org/docs} to aggregate the data from Zooniverse containing all the classifications where a candidate was identified. From the 150,000 optical and UV images, a total of 13,787 unique blue blob candidates were identified by volunteers; however, to reduce contamination of false positives classified by just 1 or 2 volunteers, we only considered blue blobs classified by at least 3 volunteers. This brought down the number of blue blob candidates to 658.  Subsequently, these classifications underwent a ranking procedure conducted by the research team (\S\ref{candidate classification}).

\subsection{Candidate classification}\label{candidate classification}

Our main goal is to find unusual and irregular blue objects in the NGVS with corresponding clumps of UV emission in the GALEX.  Many of the blue objects classified in the citizen science search often correlate to identifiable galaxies, foreground stars, or an artifact, leading us to exclude them from our candidate list. We visually differentiate false positives from BCs, which in some cases can be subjective. This led us to separate the classifications into four ranks (defined below). Three team members (M.~Jones, S.~Dey, and D.~Sand) independently assigned one of the listed ranks to each of the 658 candidates based on a visual inspection of its NGVS and GALEX images. The majority opinion was taken to assign the rank. In cases of no majority, a re-vote was conducted, with all three members debating and deciding the rank together. The ranking was performed based on the following definitions:
\begin{itemize}
    \item Rank 1 -
    These are the strongest candidates. They are faint, extended objects that appear clumpy and irregular in optical images. They appear very blue (bright in $u$-band). In addition, rank 1 BCs fulfill two key criteria: they are isolated from any galaxies within the Virgo cluster, and they have clear UV emission. The second of these emphasizes that they are centers of ongoing star formation. Figure \ref{fig:BC1} shows an NGVS and GALEX image of a rank 1 BC. Besides being irregular, it has distinct blue clumps of star-forming region in the NGVS and corresponding bright clumps of UV emission in the GALEX.

    \begin{figure*}[ht!]
        \centering
        \includegraphics[width=1\linewidth]{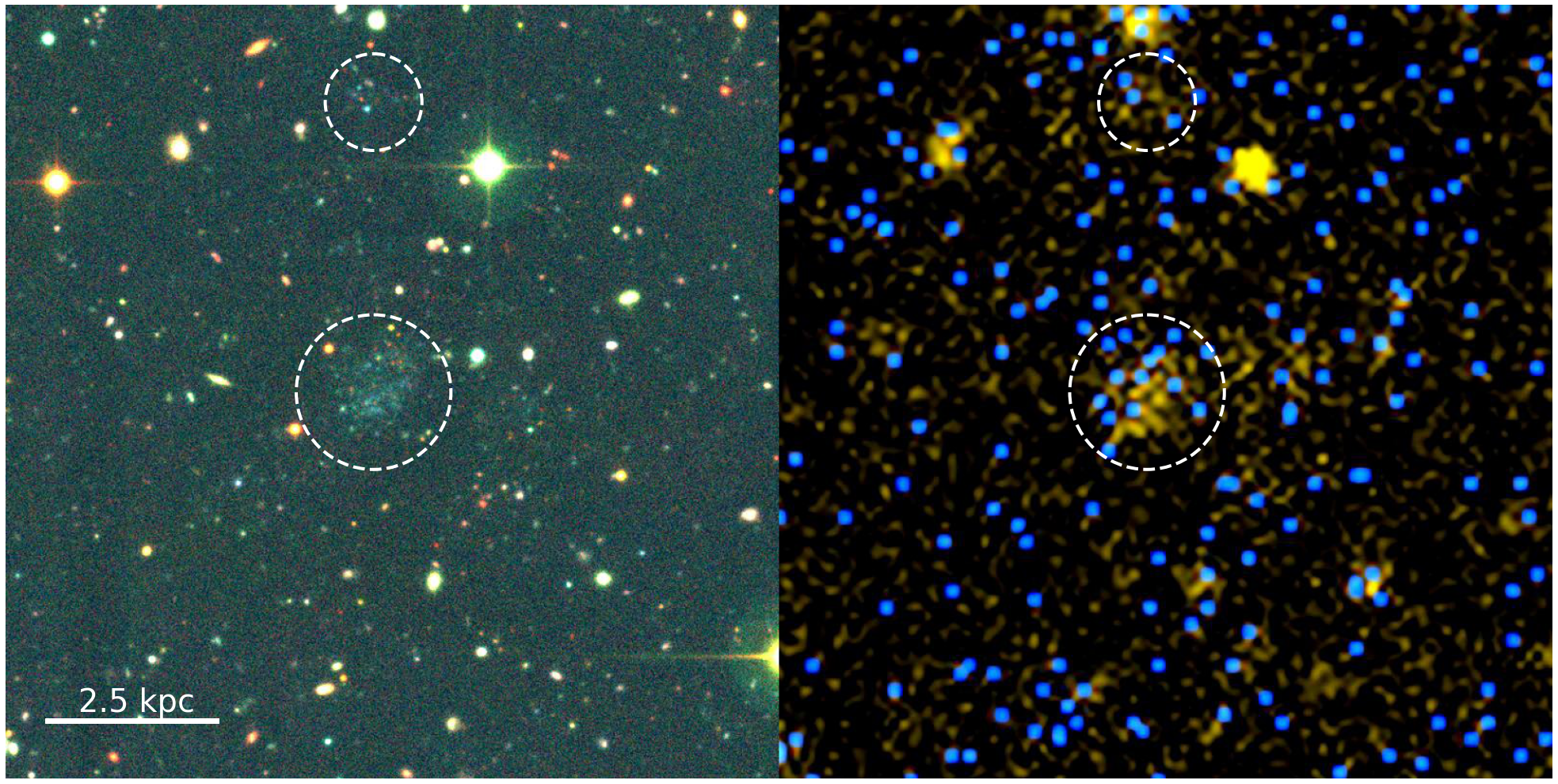}
        \caption{NGVS (left) and GALEX (right) cutouts of rank 1 candidate, BC16. There is a clear presence of blue clumps inside the white circle annotated on the NGVS cutout. It has a strong corresponding UV emission in the GALEX cutout. The GALEX cutout also has blue speckles, which are an artifact from the FUV image. This was common in the Zooniverse images in areas with poor FUV coverage. It is very irregular in shape, and using Legacy Viewer, it was confirmed that this candidate was not near any galaxy. Furthermore, it is a possible optical counterpart to a dark \hi cloud discussed in \S\ref{dark}.} 
        \label{fig:BC1}
    \end{figure*}

    \item Rank 2 -
    In comparison to rank 1 BCs, rank 2 objects are less blue and can even appear as bluish-green in the NGVS cutouts. Their structure may be either more compact than rank 1 BCs, or more smooth (less clumpy). However, they are still evident in UV emission, isolated, and irregular. We suspect some fraction of the rank 2 objects may be background galaxies. Figure \ref{fig:r2} shows a rank 2 blue blob with a bluish-green and irregular appearance and faint UV emission in GALEX.
    \begin{figure*}
        \centering
        \includegraphics[width=1\linewidth]{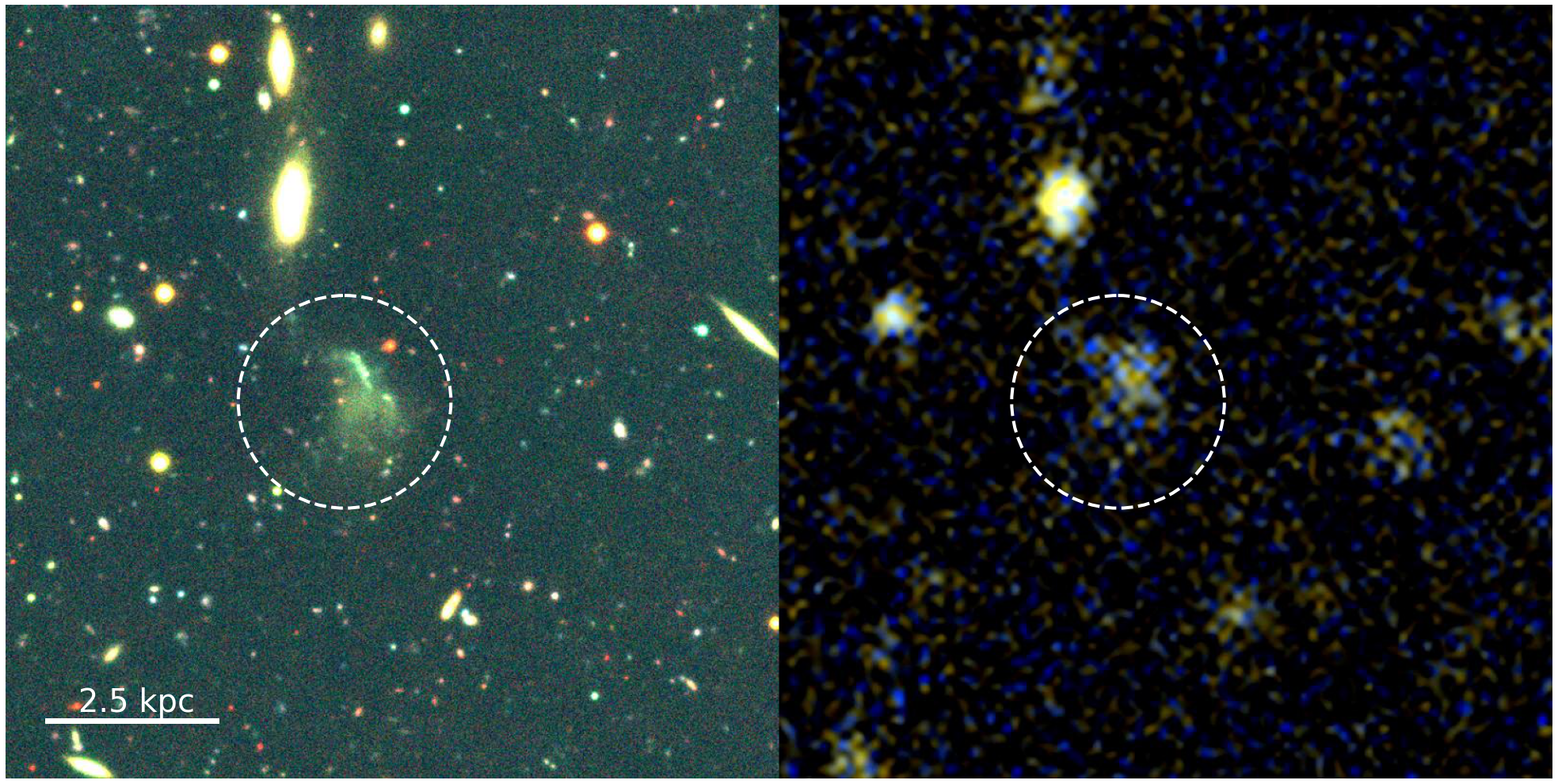}
        \caption{NGVS (left) and GALEX (right) cutouts of a rank 2 candidate, BC11. A bluish-green extended clump is present in the NGVS cutout. It has faint UV emission in the GALEX cutout. It is irregular in shape, and using Legacy Viewer, it was confirmed that this candidate was not associated with any nearby galaxy.}
        \label{fig:r2}
    \end{figure*}
    \item Rank 3 -
    These candidates have a significant likelihood of being foreground stars or background galaxies, but are not obviously so. Most of these classifications were blue, point-like sources and small diffuse objects with faint or no UV emission. We do not consider these in our analysis as we expect that they are primarily contaminants.
    
    \item Jellyfish -
    Jellyfish are structures closely related to blue blobs (\S\ref{sec:intro}), likely the result of ram pressure stripped gas forming stars in a galaxy's wake. Any blue clumps in the immediate vicinity of a galaxy were classified as jellyfish. Many objects classified as rank 1 or rank 2 BCs were subsequently reclassified as jellyfish after the initial rankings when they were inspected using large cutouts and the neighboring galaxy became apparent. 

    \item False candidates - 
    Many of the objects identified by the Zooniverse volunteers were deemed to be false candidates. These were generally foreground stars, clear background galaxies, or knots of star formation in the outer disks of spiral galaxies.

    \item Rank 0 -
    In addition to these ranks we also kept track of any unexpected object that is not a contaminant but also did not fit into the above classification scheme. We consider them ``out of pocket" candidates and designate them to rank 0. These objects are generally much larger and brighter than blue blobs. We suspect that most are dwarf galaxies being disrupted.

\end{itemize}

Out of the 658 candidates, the final ranking resulted in 19 rank 1 BCs and 21 rank 2 BCs, summarized in Table \ref{table1}. Six out of the 19 rank 1 objects were previous discoveries that were also identified by citizen scientists. Three out of the 13 new rank 1 BCs were identified by the research team prior to the citizen science search. The table lists the $i$, $g$, and $NUV$ magnitudes, stellar mass estimates, NUV star formation rate estimate, \hi mass, \hi velocity, \halpha velocity, and oxygen abundance of the candidates wherever possible. Further details of how these were measured are given in \S\ref{sec:HETdata} and \S\ref{sec:starsgas}.
Furthermore, we have compiled a catalog of all 56 jellyfish structures belonging to 44 individual galaxies identified in the Virgo cluster through our search. These are presented in Table \ref{jellytable} in Appendix \ref{jellystablesec}. 

The objects listed in Table~\ref{table1} have unique identifiers based on their coordinates (column one), however, we also list other names (column two) that have been used for the same objects (or their counterparts at other wavelengths) in the literature. In addition, this column includes a shorter BC number that was used during our earlier working catalog. We retain these as some of our earlier works and observations used these identifiers. They are also used in some figures and cases of repetitive instances in sections where labeling with the coordinate string would be impractical.

Throughout the paper, we endeavor to use the term ``blue blob" to refer to the class of objects, and ``confirmed blue blobs" for rank 1 objects that have a velocity measurement consistent with the Virgo cluster membership and a measurement confirming their high metallicity. ``Blue blob candidate" or ``BC" can refer to any of our candidates, but is mostly used for those that have not been confirmed as part of the Virgo cluster or those that have no metallicity measurement.

\subsection{Isolation criteria}\label{iso criteria}

A main differentiating factor between rank 1 and 2 blue blob and jellyfish candidates is the proximity to a potential parent galaxy. Hence, it is important to establish an isolation criteria that can distinguish between the two groups.
We search for any Extended Virgo Cluster Catalog galaxies \citep[EVCC;][]{EVCC} within 50~kpc (in projection) that are also detected in \hi \citep{ALFALFAcat}, as we assume any parent galaxy must contain \hi gas.
Furthermore, we require that the galaxy must fall in the stellar mass range of $8.3 \lesssim \log M_\ast/\mathrm{M_\odot} \lesssim 10.1$ to be consistent with the metallicity (via the mass--metallicity relation) of previously discovered blue blobs \citep{Jones2022}. 
As all candidates have already been inspected visually and appear to be quite isolated, we do not drop candidates that fail the above isolation test but rather assign a $\ddagger$ symbol to them in Table \ref{table1} to indicate that they may not be entirely isolated.

\section{Follow-up optical spectroscopy}\label{sec:HETdata}

As new rank 1 candidates were identified, they were targeted for optical spectroscopy with the Blue spectrograph of the Low Resolution Spectrograph 2 \citep[LRS2-B;][]{LRS2} on the 10m Hobby-Eberly Telescope \citep[HET;][]{HET1,HET2} at McDonald Observatory. This is an ongoing process, and to date, eight have been fully observed \footnote{BC1212+1232 (R.A. = 12:12:08, DEC. = +12:32:08) a previously rank 2 candidate was observed revealing an \halpha emission line at $>$20,000 \kms. This candidate was thus discarded as it is clearly in the background of the Virgo cluster.}. 

LRS2-B has an integral field unit (IFU) with 0.6\arcsec \ fibers and a field of view of 6\arcsec$\times$12\arcsec. In each case, the brightest clumps of the BC were centered in the IFU. The raw LRS2 data were first processed using the software \texttt{Panacea}\footnote{\url{https://github.com/grzeimann/Panacea}}, which performs bias subtraction, dark subtraction, fiber tracing, fiber wavelength calibration, fiber extraction, fiber-to-fiber normalization, source detection, source extraction, and flux calibration for each channel. Absolute flux calibration is based on standard response curves, mirror illumination measurements, and exposure throughput estimates from guider images. We then used \texttt{LRS2Multi}\footnote{\url{https://github.com/grzeimann/LRS2Multi}} to further process the fiber spectra, including background and sky subtraction, source detection on the H$\alpha$ emission line, source extraction within a 1.5\arcsec \ radius aperture, and the combination of multiple exposures.

In addition, the spectra of the six objects with detected H$\alpha$ emission are shown in Appendix~\ref{sec:HETspectra}.

\section{Stellar and gas properties}\label{sec:starsgas}

\subsection{Stellar mass estimates}\label{stellar mass}

We use the \texttt{Aperture Photometry Tool}\footnote{https://www.aperturephotometry.org/} \citep[APT,][]{APT} to find the apparent magnitude of the rank 1 and rank 2 BCs in NGVS $g$ and $i$ images. Elliptical apertures were manually constructed around each component of each BC. The sky-background estimation was done using the ``Model B" sky-algorithm of APT. This model involves marking a sky annulus around the candidate. The model then uses the sky median subtraction to give us the apparent magnitude of the source along with its uncertainty. APT is better suited for compact sources, so we split extended BCs into smaller clumps to calculate their apparent magnitudes individually and then sum them. This method is also advantageous as many candidates consist of multiple components. We also mask any contamination from clear background galaxies and stars within the aperture while doing the aperture photometry.

The apparent magnitudes were corrected for the Galactic foreground extinction using the dust reddening estimate available through the Infra-Red Science Archive (IRSA)\footnote{https://irsa.ipac.caltech.edu/applications/DUST/}, and the conversion,
\begin{equation} \label{extinction law}
    A_X = R_X \cdot E(B-V)
    \end{equation}
where $R_g$ = 3.303 and $R_i$ = 1.698 are used for the reddening coefficient of the SDSS $g$ and $i$ filters, respectively \citep[table 6,][]{Schlafly}.

We used these apparent magnitudes to estimate the stellar mass of the BCs using two scaling relations from \citet{Zibettis} (hereafter \citetalias{Zibettis}) and \citet{Taylor+2011} (hereafter \citetalias{Taylor+2011}). Both methods use color and magnitude to estimate the stellar mass-to-light ratio and, from there, a stellar mass (given an assumed distance). Equation $8$ of \citetalias{Taylor+2011} is fitted specifically for the $g$ and $i$ filters, whereas equations in \citetalias{Zibettis} can be used for many pairs of filters. In both cases, we use the absolute magnitude of the candidate in the $i$-band and the $g-i$ color. We assume a distance of 16.5 Mpc \citep{Mei+2007} for all the BCs to calculate the absolute magnitudes.  

For low-mass galaxies, the \citetalias{Zibettis} and \citetalias{Taylor+2011} methods usually bracket the actual stellar masses (refer to figure 13 of \citetalias{Taylor+2011} where they compare their result with various other stellar mass estimators). For this reason, we take the mean of the two estimates as our stellar mass estimate and their difference as the associated uncertainty for each BC. We provide the Galactic foreground extinction corrected $i$ and $g$ apparent magnitudes, as well as the stellar mass of the rank 1 and 2 blue blobs in Table~\ref{table1}. 

\subsection{Star formation rate estimates} \label{sfr}

We followed the same procedure using aperture photometry to calculate the star formation rate (SFR) from NUV fluxes as in \cite{Jones2022}. 
We restrict our SFR calculations to the NUV due to limitations in the \halpha and FUV data. We do not estimate the \halpha SFR because the HET IFU does not cover all the blue blobs completely. Additionally, FUV data is not available for all the blue blobs.
For each object, we selected the archival NUV data from GALEX with the longest exposure time. In brief, the process involves measuring flux within background-subtracted elliptical apertures and calculating uncertainties by placing 10,000 circular apertures of the same area across the GALEX tile, excluding the brightest 1\% of pixels. Magnitudes are derived using the conversions of \citet{Morrissey+2007} and corrected for extinction as in \citet{Wyder+2007}. These magnitudes are then converted to SFRs following \citet{Iglesias-Paramo+2006}. The NUV SFR rate for rank 1 and 2 BCs is provided in Table~\ref{table1}. The GALEX tiles that we used have been compiled and are available at \dataset[10.17909/2e9c-ec27]{\doi{10.17909/2e9c-ec27}}.

\subsection{Neutral gas mass}\label{HI mass}

The Arecibo Legacy Fast ALFA (Arecibo L-band Feed Array) survey, or ALFALFA \citep{ALFALFA}, was a low-resolution drift scan survey of the sky to map the \hi content of nearby galaxies ($z<0.06$) using the Arecibo Observatory. It detected \hi in more than 30,000 low redshift galaxies. The survey also covered all of the Virgo cluster. If detected, \hi spectra of blue blobs provide information about their radial velocity and \hi mass, or an upper limit on the \hi mass if undetected.

Using the ALFALFA data cubes, we extracted \hi spectra for our rank 1 and 2 BCs. Each spatial pixel in the ALFALFA data spans 1\arcmin. To extract spectra, we summed the intensity values of all the pixels within a square of length of 7 pixels centered on the pixel containing the target. The sum is then normalized by the beam area. 
We calculated the root mean square (rms) noise ($\sigma_\mathrm{rms}$) from the spectra, masking the emission due to the Milky Way and any artifacts present in the data. We designate BCs as having a detection and containing neutral hydrogen if they exhibit a distinct \hi emission line with an intensity $\geq$5$\times$rms. 

We expect any large neutral hydrogen reservoirs associated with blue blobs in the Virgo cluster to have already been detected, as this is a well-surveyed region.
Therefore, in addition to extracting our own spectra from the ALFALFA data cubes, we also match the spatial positions of our BCs with objects in existing \hi catalogs of the Virgo cluster, specifically the Arecibo Galaxy Environmental Survey \citep[AGES,][]{RhysTaylorAGES} and the ALFALFA extragalactic \hi source catalog \citep{ALFALFAcat}. We adopt the \hi mass of any \hi source that corresponds to the spatial position of the blue blob. To do this, we search for \hi sources within 2\arcmin~of each blue blob and then confirm with the NASA/IPAC extragalactic database (NED)\footnote{https://ned.ipac.caltech.edu/} whether the \hi source corresponds to the blue blob. If there is a match in both catalogs, we take the \hi mass given in the AGES catalog as this is the deeper of the two surveys. 

For blue blobs with no \hi detection and no match, we estimate upper limits for their \hi mass. To do this, we calculate the upper limit on the flux of the blue blob using the equation,
\begin{equation}\label{int flux} 
   \int S(V) \, dV <  5  \sigma_\mathrm{rms}  \sqrt{\Delta c  \Delta v}
\end{equation}
where $\Delta c$ is the channel width of 5\,\kms and $\Delta v$ is the assumed velocity width of 30\,\kms, which is typical for low-mass objects.
We then use this integrated flux (Equation~\ref{int flux}) to obtain the upper limit on \hi mass:
\begin{equation} \label{HI mass calc}
    \frac{M_{\text{HI}}}{M_{\odot}} = 2.356 \times 10^5 D_{\mathrm{Mpc}}^2 \int S(V) \, dV
\end{equation}
where $D_{\mathrm{Mpc}}$ is the distance of the blue blobs (in Mpc) assumed to be within the Virgo cluster at 16.5~Mpc.
There are 12 blue blobs matched with \hi sources in either catalog. The names and mass values of the corresponding \hi sources, along with the \hi mass limits for the remaining blue blobs, are listed in Table~\ref{table1}.

\begin{table*}[ht]
\hspace{-0.9in}
\caption{All rank 1 and rank 2 blue blobs found in the Virgo cluster (\S\ref{candidate classification}). The top portion of the table shows previously known blue blobs, the middle section shows new rank 1 candidates, and the bottom section shows rank 2 candidates. The columns are as follows: (1) Assigned name; (2) referred name within this paper and name in a major catalog (if present); (3) rank of the blue blob; (4), (5) spatial coordinates in J2000; (6), (7), (8) apparent magnitude in I, G and NUV band respectively; (9) stellar mass estimate (\S\ref{stellar mass}); (10) NUV-based SFR estimate (\S\ref{sfr}); (11) \hi mass or 5$\sigma$ upper limit; (12) heliocentric velocity of \hi emission; (13) mean velocity of \halpha clump detected with HET; (14) oxygen abundance and uncertainty (standard deviation of clumps and scatter in O3N2 calibration).}
\hspace{-0.9in}
\resizebox{1.125\textwidth}{!}{%
\begin{tabular}{ccccccccccccccc}
\toprule
Name  &    Other Name      & Rank & R.A.       & Dec.      & $m_i$ & $m_g$ & $m_{NUV}$  & $\log M_\ast$ & $\log \mathrm{SFR_{NUV}}$ & $\log M_\mathrm{HI}$ & $v_\mathrm{HI}$ & $v_\mathrm{H\alpha}$ & {[}O/H{]}& Reference \\
&    &  &        &     &   &   &   & $({M_{\odot}})$ & $({M_\odot\,yr^{-1}})$ & $({M_\odot})$ & (\kms) & (\kms) & & \\
\midrule
Previous Discoveries\\
\hline
BC1222+1328* & SECCO 1, AGC226067    & 1    & 12:21:54.1 & +13:27:37 & 20.42 $\pm$ 0.20    & 20.06 $\pm$ 0.04   &  20.36 $\pm$ 0.08              & 4.97 $\pm$ 0.40    & -3.14 $\pm$ 0.03       &      7.17    &  -142    & -153       &      -0.31 $\pm$ 0.11 & \citet{Adams+2013}\\
BC1239+1212* & BC1             & 1   & 12:39:02.8 & +12:12:16 & 20.79 $\pm$ 0.12   & 20.58 $\pm$ 0.05   & 20.58 $\pm$ 0.13 & 4.94 $\pm$ 0.35    & -3.23 $\pm$ 0.05         &     $\textless{}7.12$     &       &   1117        &      -0.34 $\pm$ 0.15  &  \citet{Jones2022} \\
BC1247+1022* & BC3, AGC226178  & 1   & 12:46:42.6 & +10:22:04 & 20.91 $\pm$ 0.20   & 20.40 $\pm$ 0.04   &    20.02 $\pm$ 0.06 & 4.73 $\pm$ 0.45    & -3.01 $\pm$ 0.02        & 7.60      & 1581  & 1584      &     -0.40 $\pm$ 0.17  & \citet{Cannon+2015}  \\
BC1226+1423* & BC4             & 1   & 12:26:25.8 & +14:23:12 & 20.41 $\pm$ 0.11   & 20.08 $\pm$ 0.02   &    20.05 $\pm$ 0.06            & 4.99 $\pm$ 0.39    & -3.02 $\pm$ 0.02        &    $\textless{}7.00$      &       &      -60     &    0.04 $\pm$ 0.15  &  \citet{Jones2022}  \\ 
BC1227+1510* & BC5  & 1    & 12:26:31.0 & +15:10:26 & 20.91 $\pm$ 0.07   & 20.77 $\pm$ 0.02   &  21.22  $\pm$ 0.06  & 4.95 $\pm$ 0.33    & -3.48 $\pm$ 0.02        &  $\textless{}7.07$        &       &      -74     &    0.01 $\pm$ 0.14  &  \citet{Jones2022}   \\
BC1230+0945* & BC6, AGC226159  & 1    & 12:29:34.5 & +09:44:31 & 21.84 $\pm$ 0.12   & 21.76 $\pm$ 0.04   &   21.28 $\pm$ 0.11             & 4.64 $\pm$ 0.31    & -3.51 $\pm$ 0.04        & 8.72      & 500   & 500       &     -0.11 $\pm$ 0.25 & \citet{JoneBC6}  \\
\hline
New Rank One\\
\hline
BC1224+1148* & BC7             & 1   & 12:24:29.5 & +11:48:19 & 23.52 $\pm$ 0.23   & 22.68 $\pm$ 0.03   & 21.65 $\pm$ 0.08 & 3.31 $\pm$ 0.56    & -3.66 $\pm$ 0.03        &     $\textless{}7.01$     &       &     137      &     -0.16 $\pm$ 0.18    \\ 
BC1247+1021*$\ddagger$ & BC12 & 1    & 12:46:36.4 & +10:21:12 & 23.11 $\pm$ 0.22   & 22.99 $\pm$ 0.06   & 22.50 $\pm$ 0.22 & 4.10 $\pm$ 0.32    & -4.00 $\pm$ 0.09         &      7.10    &   1581    &   1597        &       -0.38 $\pm$ 0.36    \\ 
BC1242+0841* & BC13            & 1   & 12:41:41.4 & +08:40:52  & 23.65 $\pm$ 0.02   & 22.75 $\pm$ 0.03   & 21.53 $\pm$ 0.13 & 3.20 $\pm$ 0.58    & -3.61 $\pm$ 0.05        &     $\textless{}7.08$     &       &      1383     &      -0.27 $\pm$ 0.37    \\
BC1243+1116 & BC15            & 1   & 12:43:10.2 & +11:15:44 & 21.85 $\pm$ 0.23   & 21.27 $\pm$ 0.04   & 22.42 $\pm$ 0.35 & 4.20 $\pm$ 0.47    & -3.97 $\pm$ 0.14        &    $\textless{}7.09$      &       &           &           \\
BC1209+1155 & BC16, AGC225880 & 1    & 12:08:48.0 & +11:54:41 & 21.45 $\pm$ 0.17   & 20.78 $\pm$ 0.03   & 20.96 $\pm$ 0.12 & 4.28 $\pm$ 0.50    & -3.38 $\pm$ 0.05         & 7.65     & 1230  &   No \halpha   &           \\
BC1236+0801*$\ddagger$ & BC17, AGESVC1 266 & 1    & 12:36:09.2 & +08:01:04  & 22.36 $\pm$ 0.20   & 21.82 $\pm$ 0.02   & 21.54 $\pm$ 0.26 & 4.03 $\pm$ 0.46    & -3.62 $\pm$ 0.10          &  7.22   & 1691  & 1671      &     -0.05 $\pm$ 0.15     \\
BC1226+1425 & BC18            & 1    & 12:26:10.0 & +14:25:14 & 22.44 $\pm$ 0.46   & 21.51 $\pm$ 0.06   & 21.59 $\pm$ 0.19 & 3.67 $\pm$ 0.59    & -3.64 $\pm$ 0.08         &     $\textless{}6.99$     &       &           &          \\
BC1230+0839* & BC25, AGESVC1 274 & 1    & 12:30:26.4 & +08:38:41  & 20.32 $\pm$ 0.03   & 20.57 $\pm$ 0.01   & 21.80 $\pm$ 0.13 & 5.52 $\pm$ 0.20    & -3.72 $\pm$ 0.05         &  6.86        &   1297    & 1311      &      -0.32 $\pm$ 0.15  \\
BC1239+1205 & BC26, AGC224219    & 1    & 12:38:32.0 & +12:04:33 & 23.95 $\pm$ 0.53   & 22.59 $\pm$ 0.05   & 22.62 $\pm$ 0.31 & 2.70 $\pm$ 0.73    & -4.05 $\pm$ 0.12        & 7.54     & 1000  & No H$\alpha$ &           \\
BC1236+1257$\ddagger$ & BC29            & 1    & 12:35:58.7 & +12:56:53  & 21.29 $\pm$ 0.19   & 21.02 $\pm$ 0.05   & 21.60 $\pm$ 0.25 & 4.70 $\pm$ 0.37    & -3.64 $\pm$ 0.10         &    $\textless{}7.11$      &       &           &           \\
BC1226+1429* & BC30            & 1    & 12:26:23.4 & +14:28:31 & 20.18 $\pm$ 0.04   & 20.27 $\pm$ 0.02   & 21.16 $\pm$ 0.08 & 5.44 $\pm$ 0.25    & -3.46 $\pm$ 0.03        &   $\textless{}6.99$       &       &     -98      &      -0.05 $\pm$ 0.15     \\
BC1218+0932 & BC31            & 1    & 12:18:23.2 & +09:31:58  & 22.89 $\pm$ 0.52   & 22.12 $\pm$ 0.08   &     21.90 $\pm$  0.36    & 3.62 $\pm$ 0.54    & -4.05 $\pm$ 0.23         &    $\textless{}7.12$      &       &           &           \\
BC1234+1121 & BC32            & 1    & 12:34:29.0 & +11:21:08 & 20.77 $\pm$ 0.05   & 20.69 $\pm$ 0.02   & 21.60 $\pm$ 0.09 & 5.07 $\pm$ 0.31    & -3.64 $\pm$ 0.04          &      $\textless{}7.10$    &       &           &           \\
\hline
Rank Two \\
\hline
BC1223+1133 & BC8             & 2    & 12:23:14.7 & +11:33:24 & 20.53 $\pm$ 0.18   & 20.62 $\pm$ 0.05  & 21.69 $\pm$ 0.22 & 5.31 $\pm$ 0.25    & -3.67 $\pm$ 0.09         &  $\textless{}7.16$        &       &           &           \\
BC1223+1110 & BC9             & 2    & 12:22:55.4 & +11:09:52 & 19.58 $\pm$ 0.04   & 20.67 $\pm$ 0.03   & 24.37 $\pm$ 1.55 & 6.55 $\pm$ 0.08    & -4.75 $\pm$ 0.62         &    $\textless{}7.09$      &       &           &           \\
BC1232+1616 & BC11            & 2    & 12:32:17.3 & +16:16:13 & 19.33 $\pm$ 0.05   & 19.69 $\pm$ 0.02   & 21.68 $\pm$ 0.14 & 6.02 $\pm$ 0.16    & -3.51 $\pm$ 0.06         &    $\textless{}7.08$      &       &           &           \\
BC1217+0850 & BC20, AGC223361 & 2    & 12:16:42.0 & +08:50:10  & 18.65 $\pm$ 0.02   & 19.10 $\pm$ 0.01   & 20.44 $\pm$ 0.19 & 6.38 $\pm$ 0.13    & -3.18 $\pm$ 0.08         & 8.27     & 1979  &           &           \\
BC1225+0755 & BC22, AGC227889 & 2    & 12:24:46.8 & +07:55:05  & 18.75 $\pm$ 0.04   & 19.06 $\pm$ 0.02   & 20.86 $\pm$ 0.18 & 6.21 $\pm$ 0.18    & -3.34 $\pm$ 0.07         & 7.39      & 801   &           &           \\
BC1242+0646 & BC27            & 2    & 12:41:32.0 & +06:46:02  & 21.39 $\pm$ 0.05   & 21.57 $\pm$ 0.03   &                & 5.04 $\pm$ 0.22    &                         &    $\textless{}7.10$      &       &           &           \\
BC1216+1235 & BC28            & 2    & 12:15:43.0 & +12:34:56 & 20.17 $\pm$ 0.03   & 20.40 $\pm$ 0.01   & 21.80 $\pm$ 0.13 & 5.57 $\pm$ 0.20    & -3.72 $\pm$ 0.05        &      $\textless{}7.09$    &       & No H$\alpha$ &           \\
BC1231+1016 & BC33, AGC223771 & 2    & 12:30:32.0 & +10:15:38 & 17.77 $\pm$ 0.01   & 18.15 $\pm$ 0.01   & 19.85 $\pm$ 0.08 & 6.67 $\pm$ 0.16    & -2.94 $\pm$ 0.03        & 7.41     & 1141  &           &           \\
BC1234+1600 & BC34            & 2    & 12:34:10.7 & +16:00:04 & 20.32 $\pm$ 0.04   & 20.69 $\pm$ 0.02   & 21.84 $\pm$ 0.16 & 5.64 $\pm$ 0.16    & -3.74 $\pm$ 0.06        &    $\textless{}7.10$      &       &           &           \\
BC1235+0816 & BC35            & 2    & 12:35:25.5 & +08:16:23.4  & 19.23 $\pm$ 0.03   & 19.75 $\pm$ 0.01   & 21.50 $\pm$ 0.17  & 6.20 $\pm$ 0.11    & -3.60 $\pm$ 0.06       &    $\textless{}7.07$      &       &           &           \\
BC1216+1145 & BC36            & 2    & 12:15:35.8 & +11:44:42 & 18.44 $\pm$ 0.02   & 18.62 $\pm$ 0.01   & 20.08 $\pm$ 0.06 & 6.22 $\pm$ 0.22    & -3.03 $\pm$ 0.03        &     $\textless{}7.05$     &       &           &           \\  
BC1229+0849 & BC37, AGC222021 & 2    & 12:28:55.4 & +08:49:01  & 19.18 $\pm$ 0.00   & 19.13 $\pm$ 0.01   & 20.15 $\pm$ 0.06 & 5.74 $\pm$ 0.30    & -3.06 $\pm$ 0.02        & 7.79    & 1306  &           &           \\
BC1218+0734 & BC38            & 2    & 12:17:43.7 & +07:34:13  & 20.84 $\pm$ 0.04   & 21.19 $\pm$ 0.02   & 22.31 $\pm$ 0.16 & 5.42 $\pm$ 0.16    & -3.92 $\pm$ 0.06         &     $\textless{}7.06$     &       &           &           \\
BC1225+1322 & BC39            & 2    & 12:24:54.9 & +13:22:21 & 20.97 $\pm$ 0.09   & 21.07 $\pm$ 0.02   & 22.78 $\pm$ 0.40 & 5.15 $\pm$ 0.25    & -4.11 $\pm$ 0.16         &    $\textless{}7.12$      &       &           &           \\ 
BC1221+1037 & BC40, AGC226131 & 2    & 12:21:13.3 & +10:37:32 & 18.79 $\pm$ 0.01   & 18.76 $\pm$ 0.01   & 19.79 $\pm$ 0.05 & 5.90 $\pm$ 0.29    & -2.91 $\pm$ 0.02        & 8.28     & 2606  &           &           \\
BC1235+0700 & BC41            & 2    & 12:35:03.7 & +07:00:17  & 20.49 $\pm$ 0.06   & 20.51 $\pm$ 0.02   & 21.33 $\pm$ 0.11 & 5.26 $\pm$ 0.28    & -3.53 $\pm$ 0.04         &   $\textless{}7.10$       &       &           &           \\ 
BC1223+0506$\ddagger$ & BC42 & 2    & 12:23:24.4 & +05:06:23  & 18.81 $\pm$ 0.02   & 19.00 $\pm$ 0.01   & 20.04 $\pm$ 0.06 & 6.09 $\pm$ 0.22    & -3.01 $\pm$ 0.03        &    9.10$^\dagger$  & 1668$^\dagger$  &           &           \\ 
BC1236+1013 & BC43            & 2    & 12:36:12.4 & +10:13:09 & 21.23 $\pm$ 0.16   & 21.59 $\pm$ 0.06   & 22.88 $\pm$ 0.41 & 5.27 $\pm$ 0.16    & -4.15 $\pm$ 0.16         &    $\textless{}7.14$      &       &           &           \\ 
BC1248+1436 & BC45            & 2    & 12:48:16.7 & +14:36:09 & 19.66 $\pm$ 0.04   & 19.64 $\pm$ 0.01   & 20.66 $\pm$ 0.08 & 5.57 $\pm$ 0.29    & -3.26 $\pm$ 0.03         &     $\textless{}7.00$     &       &           &           \\  
BC1228+1352 & BC46            & 2    & 12:28:04.6 & +13:52:04 & 20.16 $\pm$ 0.06   & 19.98 $\pm$ 0.01   & 21.01 $\pm$ 0.16 & 5.22 $\pm$ 0.34    & -3.40 $\pm$ 0.06         &    $\textless{}7.14$      &       &           &           \\  
BC1229+0907 & BC47            & 2    & 12:29:07.2 & +09:06:58  & 20.98 $\pm$ 0.02   & 21.03 $\pm$ 0.01   &      21.70 $\pm$ 0.06 & 5.10 $\pm$ 0.26    & -3.68 $\pm$ 0.02   &    $\textless{}7.08$      &       &           &           \\ \bottomrule
\end{tabular}}
\tablecomments{\\ $\ddagger$ Does not meet isolation criteria (see \S\ref{iso criteria}).\\
$\dagger$ \hi mass and velocity uncertain as the \hi counterpart may include a nearby galaxy. \\
* Confirmed blue blobs.}
\label{table1}
\end{table*}

\section{Locations and associations with known objects}
\label{sec:locations}

In this section, we present the distribution of BCs throughout the Virgo cluster and describe likely and confirmed associations between BCs and known objects. These include existing blue blobs and ``dark" gas clouds in the cluster.

\subsection{Distribution of blue blobs within Virgo}
\begin{figure*}[h!]
    \centering
    \includegraphics[width=1\linewidth]{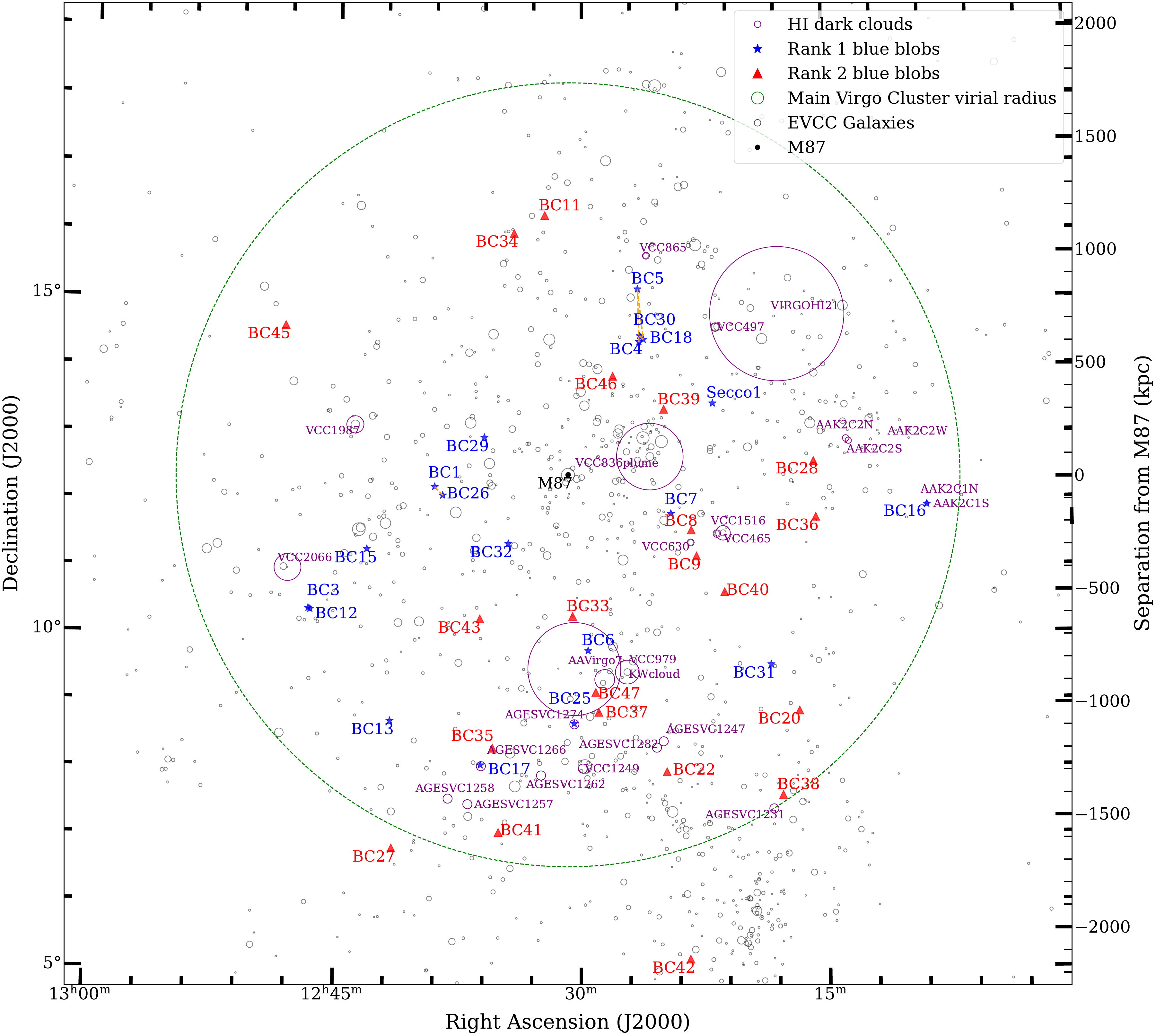}
    \caption{Spatial distribution plot of all the rank 1 (blue stars), rank 2 (red triangles), \hi dark clouds (purple circles), and Extended Virgo cluster Catalog (EVCC) galaxies (black circles) within the main Virgo cluster virial radius (green). The size of each \hi dark cloud circle is proportional to its projected size given in Table 1 of \citet{TaylorDark}. The area of each EVCC galaxy circle is determined by the total r-band flux of the represented galaxy. The virial radius of the cluster is taken to be 1.7 Mpc as per \citet{virialradius}. Blue blob candidates with potentially the same parent galaxy have been joined together with yellow dashed lines.}
    \label{fig:distBC}
\end{figure*}

Figure \ref{fig:distBC} shows the distribution of rank 1 and rank 2 BCs relative to all galaxies in the EVCC. Rank 1 BCs (including prior known BCs) are shown as blue stars, and rank 2 BCs are shown with red triangles. The figure also shows the position of all known optically dark \hi features in Virgo \citep[Table 1 of][]{TaylorDark}.  These dark clouds span a wide range of masses ($7 \lesssim \log M_\mathrm{HI}/\mathrm{M}_\odot \lesssim 9$) and can be hundreds of kpc in size (typically quite elongated). Most of these dark \hi clouds have had no optical counterparts identified to date, although the blue blob BC6 was recently associated with the ALFALFA Virgo 7 cloud complex (\citealt{JoneBC6}, and see below). In Figure~\ref{fig:distBC}, they are plotted as purple circles with radii corresponding to their angular sizes\footnote{Table 1 of \citet{TaylorDark} does not provide axial ratios or position angles of these clouds, thus we are limited to plotting them as circles even though many are known to be quite elongated.}. In the next subsection, we discuss further blue blobs that were found to be optical counterparts to some of these \ dark clouds.

\begin{figure*}
    \centering
    \includegraphics[width=1\columnwidth]{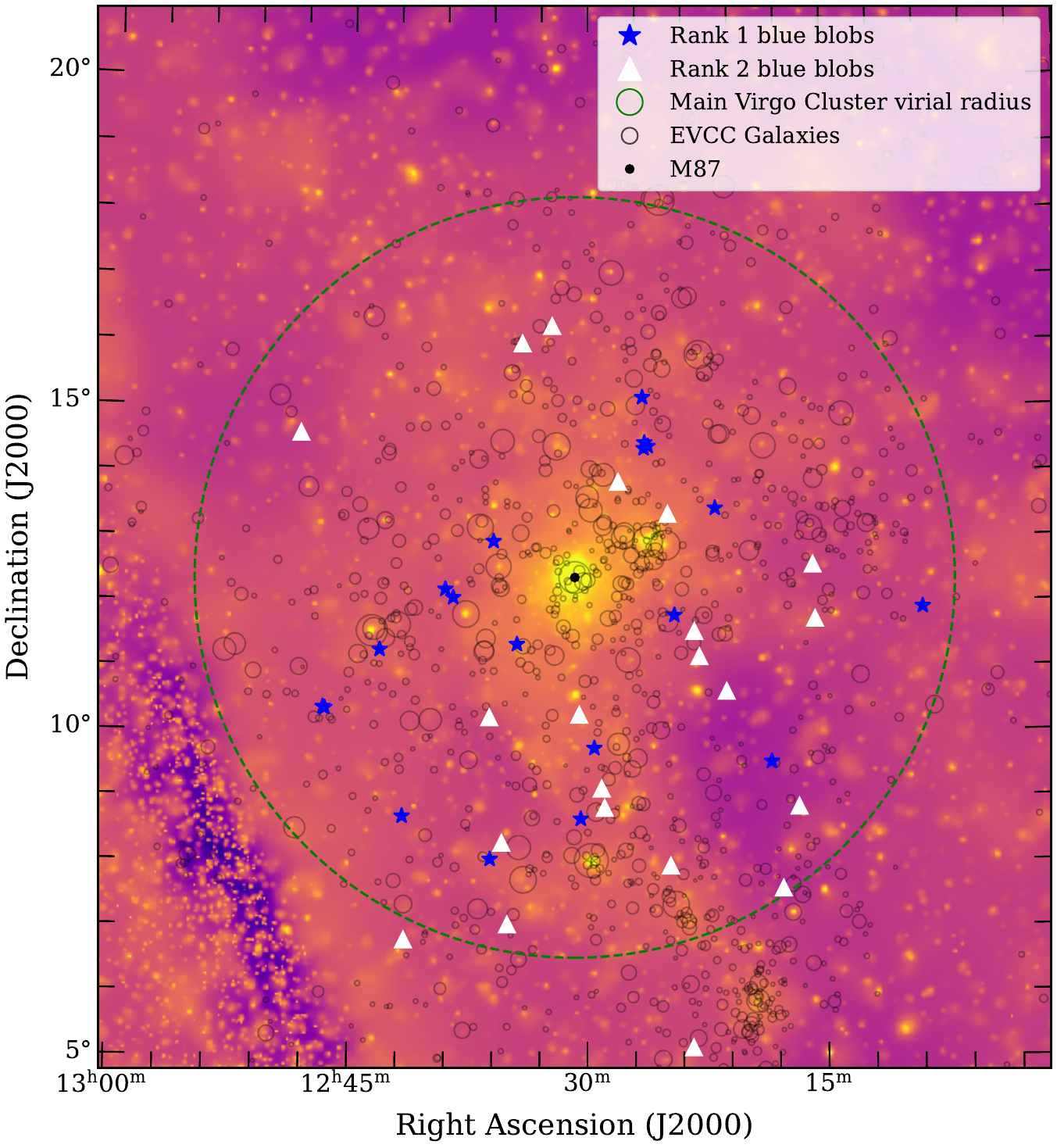}
    \includegraphics[width=1\columnwidth]{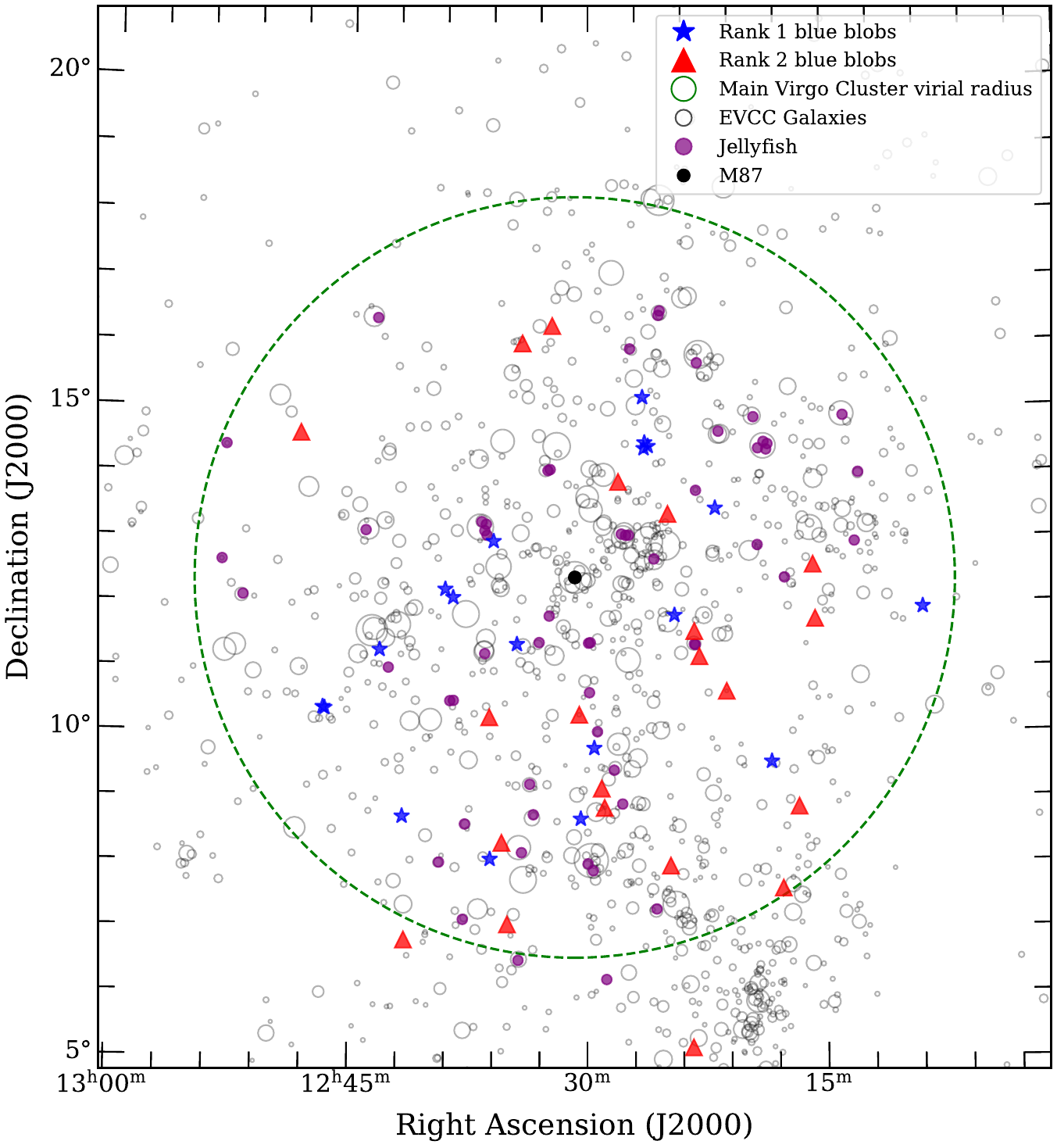}
    \caption{Left: Spatial distribution of BCs overlaid on a ROSAT mosaic of hard (0.4-2.4~keV) X-ray emission \citep{Boehringer+1994,Brown+2021} of the Virgo cluster, with yellower colors indicating the hottest and densest regions of the ICM. Galaxies \citep[from the EVCC;][]{EVCC} are shown as empty black circles. The area of each circle is proportional to the total $r$-band flux of the galaxy it represents. Rank 1 BCs are plotted as blue stars, whereas rank 2s are shown in white triangles. Right: Spatial distribution of BCs with the same markers as Figure \ref{fig:distBC} but with the addition of jellyfish structures identified in the search shown as purple dots.}
    \label{fig:xrayandjelly}
\end{figure*}

The structure of the blue blob distribution appears to be roughly similar to that of the large-scale filaments of galaxies that are falling toward the cluster center. However, there is no BC within a radius of $\sim$1$^\circ$ ($\sim$300 kpc) of the center. Rank 1 blue blob, BC1234+1121 (BC32) is the closest with a projected separation of about $\sim$400 kpc from the cluster center. The BC population starts to dwindle in regions lacking galaxies, such as in the northeast portion of the Virgo cluster imaging footprint. The left panel of Figure~\ref{fig:xrayandjelly} reinforces this point, as the BCs mostly follow areas with higher X-ray emission (which also tend to have higher galaxy density). The BCs seem to avoid the central region where the ICM is the densest and hottest. There is also a lack of BCs near and beyond the virial radius taken as 1.7\,Mpc \citep{virialradius}, although this may be due to the minimal coverage of NGVS beyond this point (cf. Figure~\ref{fig:tiles}). In the right panel of Figure~\ref{fig:xrayandjelly}, we plot the galaxies associated with the jellyfish structures identified through the search (see Table \ref{jellytable}). Except for some exceptionally isolated blue blobs, the jellyfish candidates seem to follow the same spatial pattern as the blue blobs. We will discuss more about the likely location of the parent galaxies of blue blobs in \S\ref{sec:rpsloc}.

\subsection{Blue blobs associated with HI dark clouds}\label{dark}
From our search, we have identified four BCs (all rank 1s) that appear to be the optical counterparts of previously known dark \hi clouds in the Virgo cluster. BC1230+0945 (BC6) was previously matched with one of these dark clouds by \citet{JoneBC6}, and simulations predict that these may be the sites of extragalactic star formation \citep{Ahvazi+2024}.
Where possible, we confirmed these associations by matching the H$\alpha$ velocity of the blue blobs from the follow-up observations with HET with the \hi velocity of the clouds. We discuss each of the optical counterparts below.

\subsubsection{BC1230+0945 (BC6)}
Recently, \citet{JoneBC6} confirmed the discovery of BC1230+0945 (BC6) as the stellar counterpart to the ALFALFA Virgo 7 cloud complex, which was thought to be optically dark since its discovery in 2007 \citep{Kent+2007}. BC6 lies on the NW tip of the \hi cloud complex. Its \halpha velocity of 500 \kms matches closely with the \hi velocity of 524 \kms for the cloud \citep{Kent+2009}. \citet{JoneBC6} emphasizes that its properties are consistent with those of other blue blobs \citep{Jones2022}, despite being associated with this enormous ($M_\mathrm{HI} \sim 10^9$~\Msol) structure. It has a gas fraction of $\sim$20,000 $M_\mathrm{HI}/M_\ast$ (if the whole \hi cloud complex is considered), making it the most gas-rich stellar system ever discovered.

\subsubsection{BC1236+0801 (BC17)}
\label{BC17}
The position of BC1236+0801 (BC17) overlaps with AGESVC1 266, an isolated dark \hi cloud in the southern region of the Virgo cluster. The H$\alpha$ velocity of BC17 is 1671 \kms, and the \hi velocity of the cloud is  1691 \kms, confirming it as the optical counterpart of the isolated dark \hi complex. Based on this association, the gas fraction of BC17 is $\sim$1100, with a stellar mass of $1.1\times10^4 M_\odot$ (Table~\ref{table1}).

BC17 is $\sim$6\,$\arcmin$  ($\sim$30 kpc) away from galaxy MCG+01-32-111, with a similar velocity of 1788 \kms. However, the galaxy has no obvious jellyfish structure and does not seem to have shocked gas, as seen on the Virgo Environmental Survey Tracing Ionised Gas Emission \citep[VESTIGE;][]{BoselliniVestige}. For these reasons, we believe BC17 is not a jellyfish candidate but a blue blob.

AGESVC1 266 was discussed in \citet{TaylorSnakes}, where they propose a different optical counterpart candidate $\sim$1\,$\arcmin$  ($\sim$4.7 kpc) to the south of BC17. However, this candidate was never followed up with deeper observations, and inspection of the DECaLS and NGVS imaging suggests that it is likely a background galaxy.

\subsubsection{BC1230+0839 (BC25)}
The location of BC1230+0839 (BC25) coincides with the dark \hi cloud, AGESVC1 274, also in the southern region of the cluster. 
This dark \hi complex was discussed in \citet{TaylorSnakes}, where they also propose BC25 as the optical counterpart. However, they did not obtain follow-up imaging and spectroscopy to confirm this association. Our HET spectrum reveals that the \halpha velocity of BC25 is 1311 \kms, matching the \hi velocity of the cloud (1297 \kms).
Thus, BC25 has a gas fraction of $\sim$21.8 $M_\mathrm{HI}/M_\ast$ with a stellar mass of $3.3\times10^5$ \Msol.

\subsubsection{BC1209+1155 (BC16)}
The AAK2C1S and AAK2C1N dark complexes are among the most isolated \hi clouds found in the periphery of the Virgo cluster \citep{KentDark}. 
\citet{LinDark} identified an optical counterpart to the southern cloud, AAK2C1S, which is a part of BC1209+1155 (BC16). However, we also identify additional stellar clumps overlapping with the northern cloud, AAK2C1N. 

Our follow-up observation of BC16 with HET did not detect any \halpha emission; hence, we could not match its velocity to that of the \hi clouds. However, based on its morphology and size, as seen in the NGVS images, it appears to be within the cluster and associated with these two clouds. The stellar mass of the extension in the northern cloud is $3\times10^3$~\Msol\ and has a gas fraction of $\sim$5900 if we only consider the gas in the coincident cloud (AAK2C1N). The main clump of BC16 has a stellar mass of $1.6\times10^4$ \Msol\ and a gas fraction of $\sim$1500 if only considering the gas in AAK2C1S. For the whole system, the combined stellar mass is $2\times10^4$~\Msol and has a total gas fraction of 2000 (this time, including the gas mass of both clouds). 

\subsection{Extensions of previously known blue blobs}
Four new rank 1 BCs were found in close proximity to previously known blue blobs BC1239+1212 (BC1), BC1247+1022 (BC3), BC1226+1423 (BC4), and BC1227+1510 (BC5). These new candidates are likely physically associated with the previously known blue blobs, and in some cases, we have confirmed this association through velocity and metallicity measurements. We describe each of the four cases below.

We find BC1226+1425 (BC18) and BC1226+1429 (BC30) less than 4\arcmin \ (190\,kpc) away from BC4. \citet{Jones2022} proposed the same point of origin for BC4 and BC5, and it is likely that all four of these objects originate from the same parent galaxy. BC30 shares a similar \halpha velocity (-98\,\kms) to BC4 (-60\,\kms) and BC5 (-74\,\kms). It also has comparable metallicity, $[\mathrm{O/H}] = -0.05 \pm 0.14$, while BC4 and BC5 have $[\mathrm{O/H}] = 0.04 \pm 0.15$ and $0.01 \pm 0.14$, respectively. BC18 does not have any available velocity or metallicity measurements, but based on its (projected) location (Figure~\ref{fig:distBC}), it is also likely a component of the same large structure.

BC1247+1021 (BC12) appears to be an extension of BC3, approximately 9~kpc to the southwest in projection. This small clump exhibits an \halpha velocity of 1597~\kms, closely matching BC3's \halpha velocity of 1584~\kms. Both exhibit nearly identical metallicity values, with $[\mathrm{O/H}] = -0.38 \pm 0.36$ for BC12 and $-0.40 \pm 0.17$ for BC3. BC12 lies in the same direction as VCC~2037 (relative to BC3), which \citet{Jones2022a} suggested as a possible point of origin, and is 70~kpc to the southeast and at a similar velocity (1507~\kms).

BC1239+1205 (BC26) is a possible extension of BC1. It is about 53~kpc southwest of BC1 in projection. Unfortunately, we have not yet obtained an independent velocity or metallicity measurement, so we cannot confirm this association. BC26 is also in close proximity to the ALFALFA \hi source AGC224219, which is thought to correspond to the LSB galaxy LSBVCC~79 \citep{Davies2016,ALFALFAcat}. The \hi centroid of AGC224219 is closer to LSBVCC~79 than BC26, but LSBVCC~79 lacks any GALEX emission, making it an unusual counterpart for an \hi source. If the \hi in this source has a disturbed morphology, as does the gas in the vicinity of BC3 \citep{Jones2022a}, then it is possible that the \hi centroid could be misleading and that this gas might be associated with BC26 rather than LSBVCC~79. However, without follow-up synthesis imaging of the gas and an \halpha velocity measurement for BC26, this potential association cannot presently be verified.

\section{Stellar mass, SFR, and metallicity}
\label{sec:stellarprops}
Using the methods outlined in \S\ref{sec:HETdata} \& \ref{sec:starsgas}, we determined the stellar mass, SFR, \hi mass, and metallicity of the rank 1 and 2 BCs. The results are summarized in Table \ref{table1}. 
In this section, we compare rank 1 and 2 BCs with a sample of ALFALFA galaxies, well-known extremely low-mass star-forming dwarf galaxies, and an empirical relation for the main sequence of low-mass, star-forming galaxies. We also compare the metallicity of the BCs with that of Local Group dwarfs, Local Volume dwarfs, tidal dwarf galaxies, and extremely metal-poor galaxies.

\begin{figure}
    \centering
    \includegraphics[width=1\linewidth]{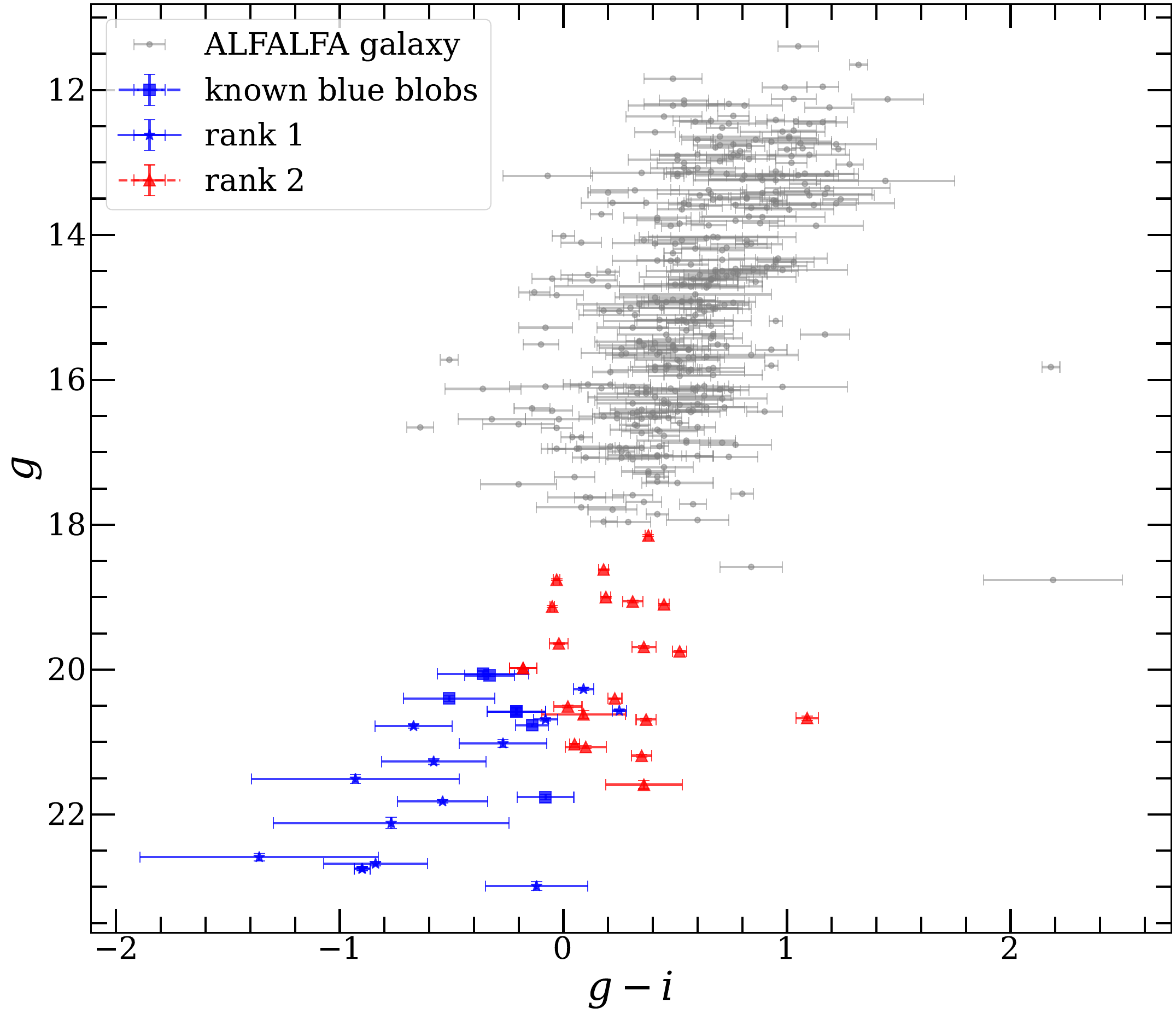}
    \caption{Color-Magnitude-Diagram of rank 1 and 2 BCs, and ALFALFA galaxies within 20\,Mpc \citep{ALFtable2}. Rank 1 BCs are shown in blue stars with blue error bars, rank 2 BCs in red triangles with red error bars, and ALFALFA galaxies in gray dots and error bars. Prior known blue blobs are shown as blue boxes with blue error bars.}
    \label{fig:CMD}
\end{figure}

\subsection{Mass, SFR, and gas-fraction relations}
Blue blobs are still a recently discovered class of objects, and their nature is not yet entirely clear.

Previously, it has been suggested that these objects are not galaxies but young and isolated star-forming clouds formed from stripped gas \citep{Sand+2017, Bellazzini+2018, Jones2022}. 
However, it is worth comparing the properties of the current, larger sample of blue blobs to ALFALFA galaxies from the ALFALFA-SDSS galaxy catalog \citep{ALFtable2}, low-mass star-forming dwarf galaxies, and the main sequence for low-mass star-forming galaxies. This might reveal whether blue blobs could be consistent with the galaxy population or if they are, in fact, not at the distance of the Virgo cluster. Or it might reinforce the notion that they are not galaxies.

In Figure \ref{fig:CMD}, we compare the $g$ and $i$ magnitudes of rank 1 and 2 BCs with ALFALFA galaxies that are within 20\,Mpc. We see that the majority of the rank 1 BCs are significantly bluer ($g\,-\,i$ $\lesssim$ 0) and fainter (g $<$ 20). However, the rank 2 BCs are less blue and brighter than the rank 1 candidates, yet much fainter than most of the ALFALFA galaxies. Rank 1 candidates are typically brighter in $g$-band than in $i$-band, and those with large uncertainties in their colors are objects that are nearly undetected in $i$-band. 

\begin{figure*}[ht!]
    \centering
    \includegraphics[width=.475\linewidth]{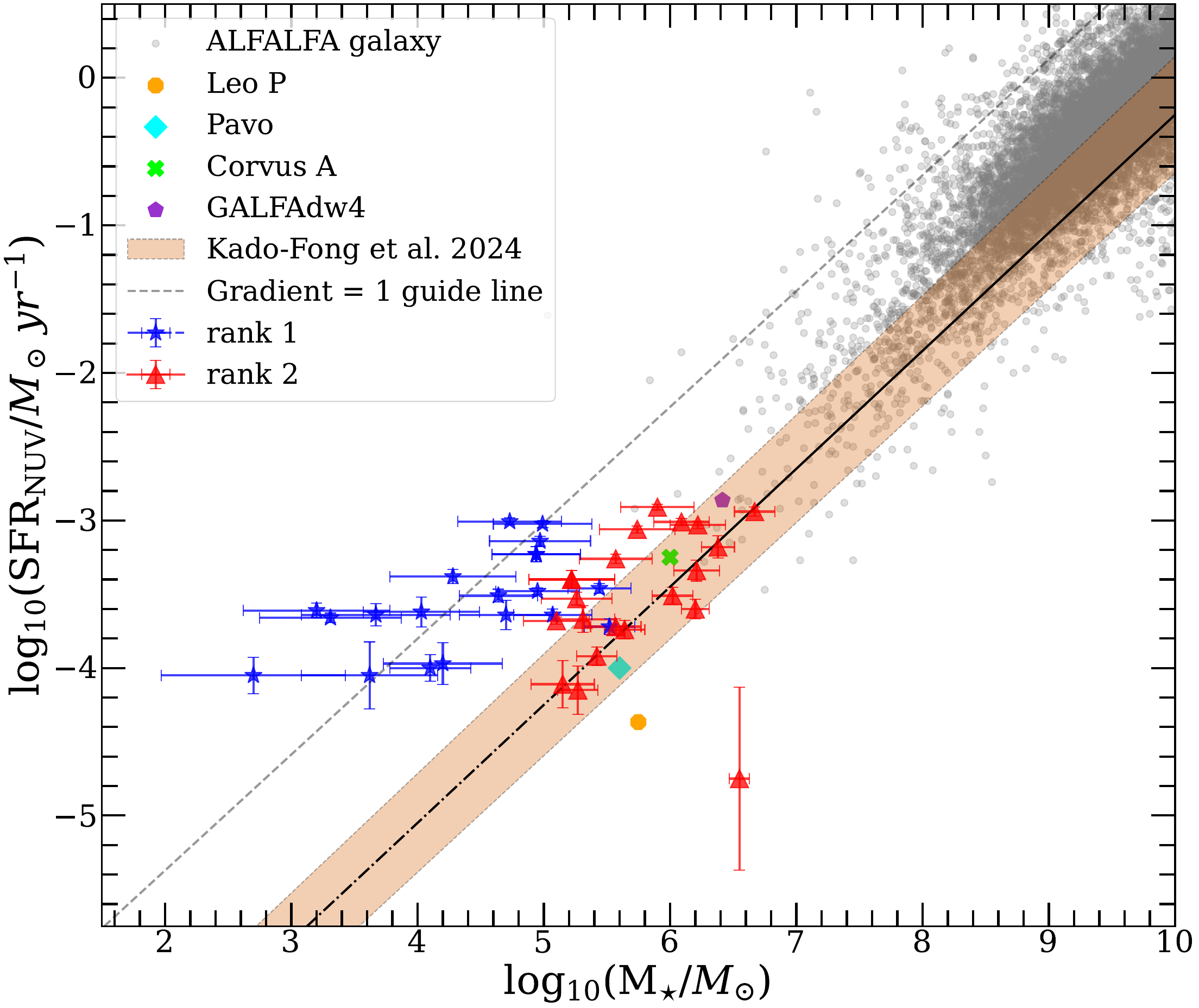}
    \includegraphics[width=.475\linewidth]{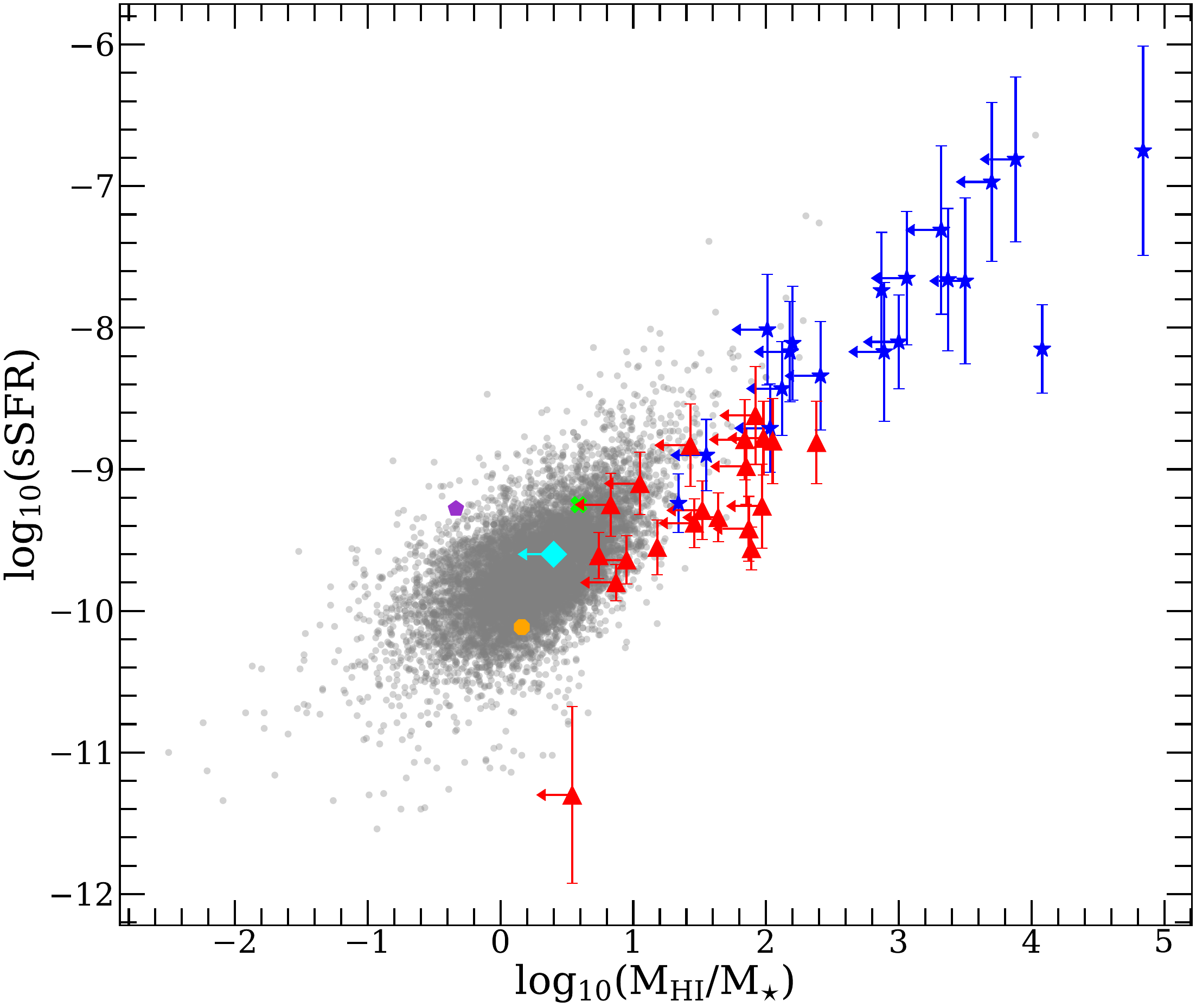}
    \caption{\textit{Left}: NUV star formation rate as a function of the stellar mass of blue blobs and comparison with low-mass star-forming dwarfs, ALFALFA galaxies, and the empirical star formation main sequence measurement.
    Rank 1 blue blobs are shown by blue stars, while rank 2 blue blobs are red triangles. The sample ALFALFA galaxies are represented by gray dots, with their stellar mass estimates calculated following \citetalias{Taylor+2011} and star formation rate calculated using NUV data. The sample of low mass, star-forming dwarf galaxies include Leo P \citep{leo, LeoP2}, GALFA-Dw4 \citep{Galfa}, Pavo \citep{Pavo}, and Corvus A \citep{Corvus} in orange, purple, cyan and lime green respectively. The empirical MS measurement is from \citet{erin-dong} in brown.
    \textit{Right}: Specific star formation rate as a function of gas-fraction ($M_{HI}/M_*$) of rank 1, rank 2, and the same comparison sample of low-mass star-forming dwarfs and ALFALFA galaxies. Symbols and data references of our sources and the sample galaxies are the same as in the left panel.
    There is an evident distinction between the three groups in both the plots.}
    \label{fig:SMvsSFR}
\end{figure*}

The left panel of Figure~\ref{fig:SMvsSFR} shows the SFR--$M_*$ relation of our sources. Besides a slight overlap, there is a clear contrast between rank 1 and 2 objects. This implies a strong correlation between our ranking criteria, which were solely based on morphology and color, and the stellar mass estimates of the candidates. However, both rank 1 and 2 objects have similar ranges of NUV SFR estimates.

The BCs have a lower stellar mass and SFR than 2500 randomly selected ALFALFA galaxies.\footnote{We select 2,500 galaxies to avoid overplotting and prevent the plot from becoming too crowded.} This is expected, considering the extremely blue and faint appearance of blue blobs. Both ranks are mostly confined within $-4 \lesssim \log{( \frac{\mathrm{SFR_{NUV}}}{\mathrm{M_{\odot}\,yr^{-1}}}} ) \lesssim -3$ and $2.5 \lesssim \log{( \frac{\mathrm{M_{*}}}{\mathrm{M_{\odot}}}} ) \lesssim 7$.

Our second comparison sample includes Leo P \citep{LeoP2,leo}, GALFA-Dw4 \citep{Galfa}, Pavo \citep{Pavo}, and Corvus A \citep{Corvus}. These objects are some of the lowest mass star-forming galaxies known and are, therefore, the most suitable to compare to blue blobs. Yet even these galaxies are more massive than most rank 1 blue blobs. The dwarf sample appears to occupy a similar parameter space as most rank 2 BCs. However, it is important to note that these galaxies are significantly closer, at distances of $\lesssim$ 4\,Mpc, compared to the assumed distance of the blue blobs, which is 16.5\,Mpc.

Next, we plot the star-forming main sequence relation of \citet{erin-dong} at $z = 0$, derived using a sample of 23258 low-mass star-forming galaxies that were background objects identified in the Satellites Around Galactic Analogs (SAGA) Survey \citep{Geha+2017,Mao+2021,Mao+2024} in the mass range of $6 \lesssim \log{( \frac{{M_{*}}}{\mathrm{M_{\odot}}}} ) \lesssim 10$. The main sequence relation does a very good job fitting our selected sample of low-mass star-forming dwarfs (e.g., Corvus A and Pavo), as well as the majority of rank 2s.

Given that both SFR and M$_*$ are distance-dependent quantities and the uncertainty surrounding the distances of the blue blobs, it is imperative to investigate how the blue blobs would be positioned on the plot if their distances differed from the assumed 16.5\,Mpc. 
An effective change in the distances will cause equal (fractional) changes in both SFR and stellar mass, as both quantities depend on the square of the distance. This can be visualized by shifting their positions parallel to the guideline shown in the gray dashed line (this line has a gradient of unity). A rank 1 blue blob of mass $\sim 10^4$\,\Msol \ (at 16.5~Mpc) would need to be at a distance of over 150\,Mpc away to have a stellar mass of $10^7$\,\Msol, similar to that of the lowest mass ALFALFA galaxies. This is not possible as the previously known blue blobs were resolved into individual stars by the Hubble Space Telescope \citep{Sand+2017,Jones2022}. 
Moreover, if the rank 1 BCs are shifted along the guideline, most will remain outliers compared to the ALFALFA galaxies and the main sequence at higher stellar masses and SFRs.  

This is not the case for rank 2s, which have smoother morphologies and could be significantly farther away. Since there is no HST imaging or H$\alpha$ velocity measurement for the rank 2 blue blobs, it is far from certain that they reside within the Virgo cluster. Rank 2 BCs seem to lie along the plane of the ALFALFA galaxies, whereas the rank 1 BCs are shifted slightly from the plane due to their lower stellar mass. This hints that rank 2s have the possibility of being background galaxies.          
Given the greater uncertainty in the distances of rank 2 BCs, it is possible that most of them have a higher stellar mass than estimated. Since \citet{erin-dong} reports an almost linear main sequence relation at $z = 0$, a shift in rank 2 BCs' position in the plot will still be along the relation and lie within its scatter range. 
They could also potentially end up within the cloud of ALFALFA galaxies.\footnote{We note that the \citet{erin-dong} relation falls below most of the ALFALFA galaxies. This is likely a result of the fact that these galaxies are \hi-selected and, therefore, represent a sample that is biased towards gas-rich galaxies that also, presumably, tend to have elevated SFRs.}

In summary, it appears plausible that at least some rank 2 BCs could be distant low-mass galaxies that have been misidentified, but this is not true of rank 1 BCs, most of which do likely reside in the Virgo cluster. Furthermore, most rank 1 BCs lie well above (higher SFR) the star-forming main sequence relation for low-mass galaxies, yet all known galaxies with a stellar mass $<10^5$\,\Msol \ are quenched ultra-faint dwarfs (UFDs). However, unlike blue blobs, the known UFDs thus far are not star-forming due to quenching by cosmic reionization in the early universe \citep{Brown2014}. Hence, blue blobs are not likely to be normal galaxies.

Finally, given the uncertainty in the distance to many BCs, we have made a distance-independent plot by comparing the specific star formation rate (SFR/M$_*$) and the gas fraction ($M_\mathrm{HI}/M_*$) shown on the right panel of Figure~\ref{fig:SMvsSFR}. Our sample of extremely low-mass star-forming galaxies occupy the same parameter space as the ALFALFA galaxies, however, the blue blobs do not. There are no galaxies with as high a gas fraction as the BCs at the furthest end of the plot. This makes them some of the most gas-rich stellar systems ever discovered. It is not currently clear why all the BCs lie along a continuation of the relation of the galaxies, however, this is likely a result of our implicit detection sensitivities. 
It is also important to note that the majority of the BCs have an upper limit on the \hi mass (see \S\ref{HI mass}). 

This means that they could have gas fractions that are significantly lower (to the left) than where they are plotted.
Most rank 2 BCs display a comparable specific star formation rate (SFR) to those of the ALFALFA galaxies and low-mass, star-forming dwarf galaxies. A leftward shift could align rank 2s with the population of galaxies.  However, the rank 1 BCs with upper limits will lie well above the cloud of ALFALFA galaxies even if shifted to a considerably lower gas fraction. Furthermore, some of the rank 1 BCs with the most extreme gas fraction are \hi detections, not upper limits. Thus, again we see that rank 1 BCs are inconsistent with the properties of galaxies even in a distance-independent parameter space.

\subsection{Metallicities of newly confirmed blue blobs}\label{metallicity discussion}
\begin{figure}[h!]
    \centering
    \includegraphics[width=1\linewidth]{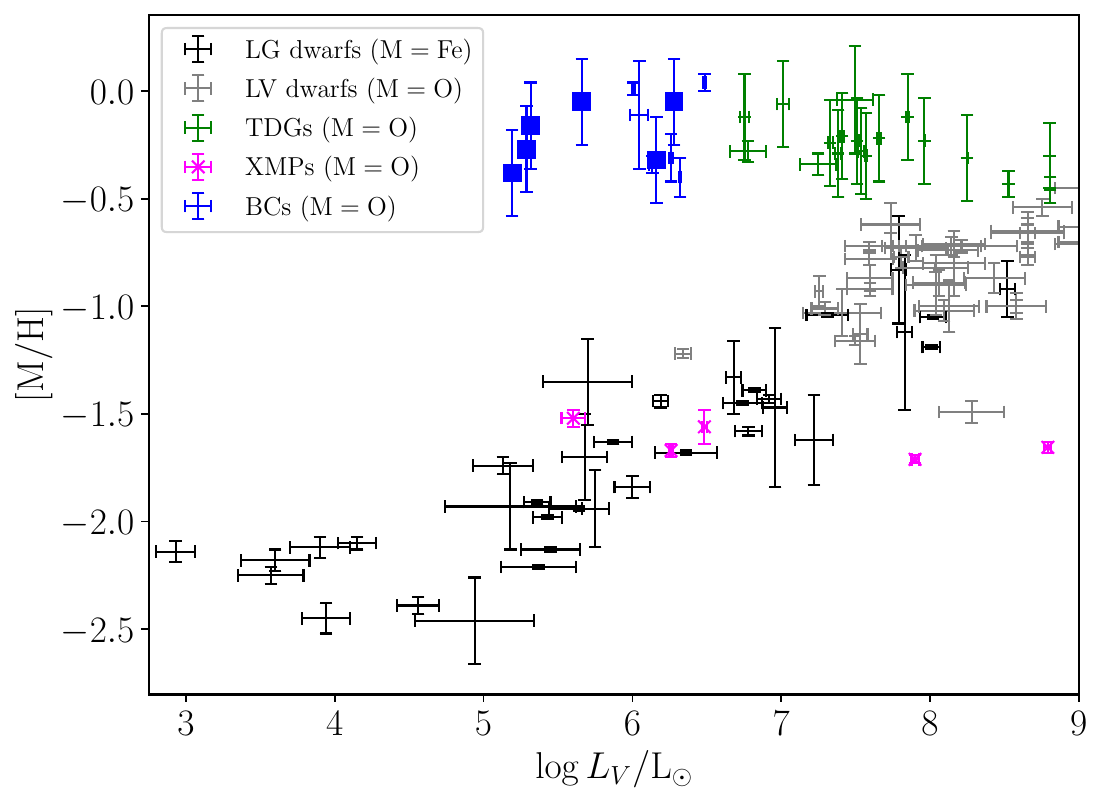}
    \caption{V-band luminosity versus metallicity (relative to solar) for BCs, Local Group dwarfs \citep{Kirby+2013}, Local Volume dwarfs \citep{Berg+2012}, TDGs \citep{Duc+1998,Weilbacher+2003,Duc+2007,Croxall+2009,Lee-Waddell+2018}, and extremely metal-poor galaxies \citep[XMPs,][]{Skillman+2013,leo,Hirschauer+2016,Hsyu+2017,Izotov+2019,McQuinn+2020}. The new blue blobs are shown by blue squares with blue error bars.}
    \label{metallicity_plot}
\end{figure}

Figure 11 of \citet{Jones2022} compares the metallicity of blue blobs with other stellar systems of similar luminosity, such as Local Group and Local Volume dwarfs, tidal dwarf galaxies (TDGs), and extremely metal-poor galaxies (XMPs). In Figure~\ref{metallicity_plot}, we reproduce this plot, but now with double the sample size of confirmed blue blobs with metallicity measurements. The newly confirmed blue blobs (blue square markers) follow the previous trend and exhibit higher metallicities ($-0.5 \lesssim \mathrm{[M/H]} \lesssim 0$) compared to local dwarf galaxies.  Their metallicities are similar to TDGs, but they are less luminous and massive. As \citet{Jones2022} argues, for such low-mass stellar systems to have these high metallicities, a viable formation mechanism needs to be capable of explaining the pre-enrichment of their gas. Gas stripping from a large parent galaxy is a natural explanation for these properties. Furthermore, since the new BCs fall within the metallicity range of previously known BCs, it suggests that they also likely originate from parent galaxies within a similar mass range, $8.3 \lesssim \log M_\ast/\mathrm{M_\odot} \lesssim 10.1$ \citep{Jones2022}.

\section{Phase space locations of blue blobs} \label{sec:rpsloc}

\begin{figure}[h!]
    \centering
    \includegraphics[width=1\linewidth]{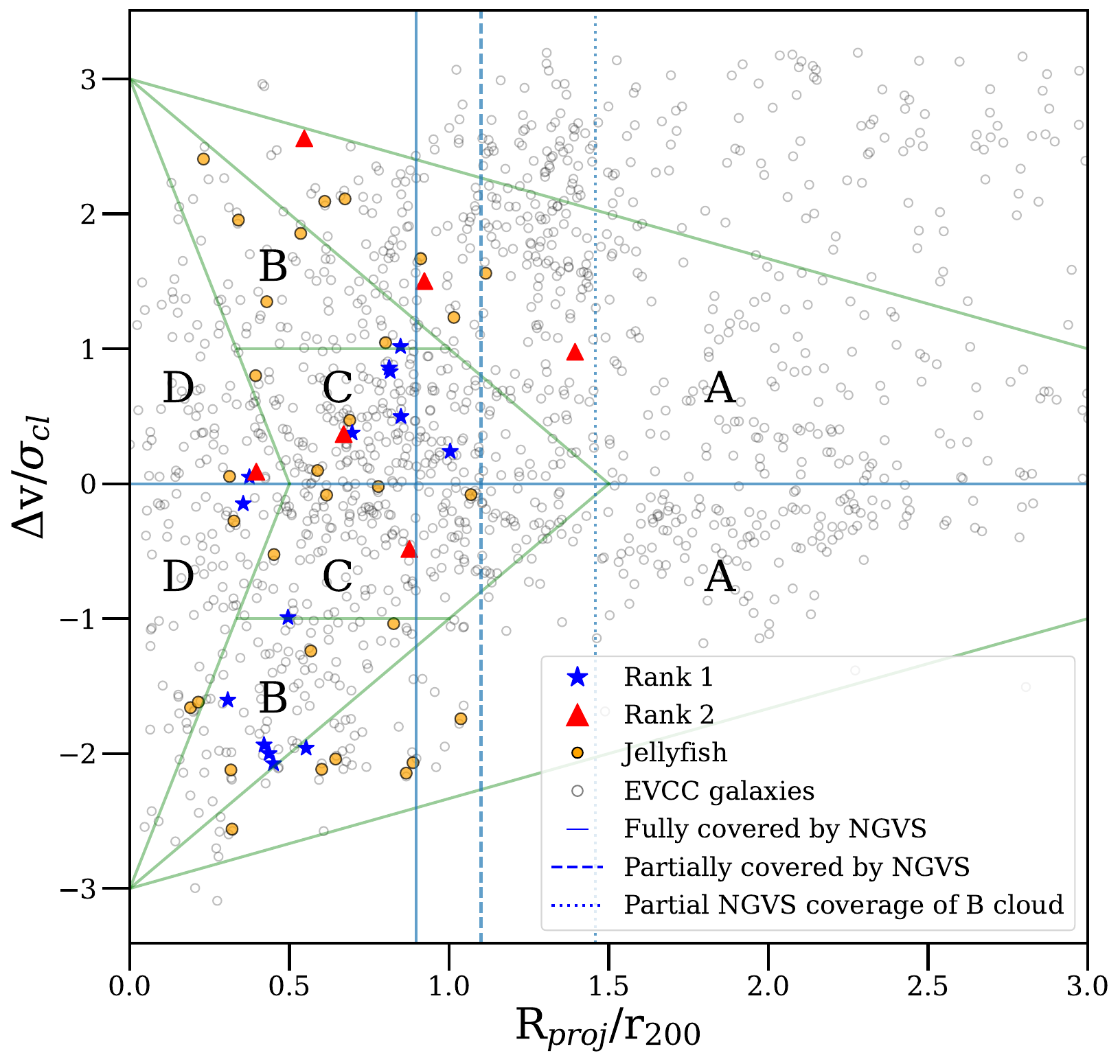}
    \caption{Projected phase-space diagram of blue blobs, jellyfish galaxies, and EVCC galaxies. Rank 1 and rank 2 objects are shown by blue stars and red triangles, respectively. Jellyfish candidates are orange dots, and EVCC galaxies are black circles. We use the same normalizing constants; cluster velocity dispersion ($\sigma_{cl}$) and virial radius ($r_{200}$) as \citet{mun}. We follow the labeling of regions based on the time since the infall of galaxies into the cluster, as shown in \citet{mun}. The projected radius fully covered by NGVS is marked by a solid vertical line. Partial NGVS coverage extending to the Virgo virial radius is shown with blue dashed lines. Part of NGVS coverage extending out to the Virgo B cloud is shown as a dotted line.
    }
    \label{fig:RPS}
\end{figure}

\citet{Jones2022} presented evidence favoring ram pressure stripping as the most viable mechanism for forming blue blobs and discussed potential parent galaxies for a few systems. With our current, considerably larger sample of BCs, we are now able to consider the question of their points of origin in a more statistical manner, rather than on a case-by-case basis.

Figure 2 of \citet{mun} labels the orbital stages of a cluster member galaxy depending on its location in a projected cluster-centric radius versus velocity phase space diagram. Figure \ref{fig:RPS} shows this projected phase-space diagram for EVCC galaxies, BCs, and jellyfish galaxies with M87 taken as the center of the cluster and the mean radial velocity of 1088 \kms of the M87 subgroup \citep{mun} taken as the systemic velocity of the cluster. Regions A, B, C, and D in the figure represent different categories of member galaxies designed to approximately trace the time since infall into the cluster. Region A includes ``first infallers," galaxies entering the virial radius of the cluster for the first time. Region B consists of ``recent infallers," galaxies that fell into the cluster within the last 0 to 3.63 Gyr. Region C comprises ``intermediate infallers," which fell into the cluster between 3.63 and 6.45 Gyr ago. Finally, Region D contains ``ancient infallers," galaxies that entered the cluster more than 6.45 Gyr ago \citep{rhee}. While a projected phase-space diagram like Figure \ref{fig:RPS} only roughly estimates the time since infall for objects, a more accurate estimate can be determined using a three-dimensional phase-space diagram. However, this would require knowing the logarithmic radial distance of the objects from the cluster's center and their tangential velocity. Thus, Figure \ref{fig:RPS} is only valid for statistical interpretation and not for individual objects. Although blue blobs are not galaxies and, therefore, strictly speaking, these phase space regions are not applicable to them, their location within the space may provide clues regarding their parent galaxies. 

Our BCs are mostly coincident with galaxies in the recent or intermediate infall regions and mostly avoid the ancient infall region. Taken at face value, this suggests that the parent galaxies of blue blobs may have been cluster members for some time rather than on their first infall. 
However, while ram pressure will do little to decelerate the infall of a parent galaxy (as most of the mass is in dark matter and stars), initially, blue blobs are presumably made entirely of gas and should come to rest with respect to the ICM relatively rapidly. Thus, blue blobs could potentially originate from galaxies in the first infall regions but `sink' in the phase space towards the intermediate regions. 

Almost no BCs were identified in region A (first infall), which would be a natural result of either scenario described above. However, it should also be noted that much of this region is missing coverage in NGVS and was, therefore, not searched in the Zooniverse project. The lack of BCs in the portions of region A with coverage could also indicate that ram pressure stripping in this part of the cluster is not strong enough to produce blue blobs. Additionally, in contrast to high mass clusters, large galaxies in the Virgo cluster are not completely stripped in one orbital pass \citep[e.g.][]{chung+2009}, allowing stripped material to be generated over several orbits. Galaxies associated with jellyfish candidates and BCs mostly share the same parameter space; however, jellyfish candidates are slightly more concentrated in the recent infall regions. 

In Appendix~\ref{sec:phase_map}, we re-plot Figure~\ref{fig:distBC} but now coloring each EVCC galaxy based on its phase space category. BCs also tend to be co-spatial with galaxies in regions B and C of the phase space. Assuming that the separations between blue blobs and their parent galaxies are typically not enormous (e.g., $\lesssim$2$^\circ$ in projection), then this supports the interpretation that the typical parent galaxies are genuinely those residing in the recent/intermediate regions of the infall phase space. 

One possible way to estimate the distance between a blue blob and its parent galaxy would be to determine how long ago the blue blob began forming stars, which would set a minimum time that it has been separated from its parent object. To determine this will require deep imaging of their resolved stellar populations which would allow their star formation histories to be measured. However, imaging at the requisite depth is only feasible with JWST.

\section{Conclusions}
\label{sec:conclusions}

Through a citizen science search covering the entire Virgo cluster, we have identified 34 new blue blob candidates in NGVS and GALEX imaging and confirmed 6 of these with emission line spectroscopy with the HET. We separate the blue blob candidates into rank 1 and rank 2 objects, primarily based on their appearance in the optical and UV images (see \S\ref{candidate classification}). We find that rank 1 blue blob candidates are broadly consistent with previously discovered blue blobs \citep{Jones2022} in terms of their blue colors ($g-i <$ 0), strong UV emission, clumpy morphologies, low stellar mass estimates ($2.5 \lesssim \log{( \frac{M_{*}}{\mathrm{M_{\odot}}}} ) \lesssim 5.5$), low SFR ($-4 \lesssim \log{( \frac{\mathrm{SFR_{NUV}}}{\mathrm{M_{\odot}\,yr^{-1}}}} ) \lesssim -3$), and isolation. The newly confirmed blue blobs also have velocities consistent with Virgo cluster membership, as well as similarly high metallicities ($0 \lesssim \mathrm{[O/H]} \lesssim -0.5$) to previously identified blue blobs \citep{Beccari+2017,Bellazzini+2022,JoneBC6} and TDGs.

Rank 2 BCs have evident UV emission and similar SFR estimates but are less blue and have smoother morphologies and higher stellar mass estimates than rank 1 BCs.
To date we have only observed one rank 2 BC with optical spectroscopy, but no emission lines were detected. Thus, at present, there are no metallicity measurements for rank 2 BCs, and the only velocity measurements come from those that are coincident with known \hi sources.

We compared the properties of blue blobs with a sample of low-mass star-forming dwarfs and empirical relations of the star-forming main sequence of galaxies. The combination of stellar masses, SFRs, and metallicities of rank 1 BCs is inconsistent with the galaxies in these comparison samples. A few rank 1 BCs are also amongst the most gas-rich stellar systems ever found, with gas fractions ($M_\mathrm{HI}/\mathrm{M_\odot}$) over 1000. Given these properties, we come to the same conclusion as previous works, that blue blobs are likely isolated star-forming clouds in the Virgo cluster formed as a result of extreme ram-pressure stripping episodes. However, some rank 2 BCs may be false positive candidates and could be galaxies in the background of the cluster.

We find a lack of blue blobs in the cluster center and areas with low galaxy density. However, they appear to be more concentrated along the filamentary structures building the cluster. We also confirm the association of three new blue blobs with previously known ``dark" \hi clouds in the cluster \citep{TaylorDark,KentDark}. 
Based on the positions of blue blobs and jellyfish structures in projected phase space (Figure~\ref{fig:RPS}), we conclude that the parent galaxies of blue blobs have likely been in the cluster for intermediate periods (3.63 to 6.45 Gyr) and are probably not on their first infall. 

Virgo was likely the first place that these objects were identified due to its proximity and the large number of surveys that cover it. The Fornax cluster is the next nearest cluster, and given its proximity and the depth of the Fornax Deep Survey \citep[FDS;][]{FDS}, it appeared to be the next most viable location for identifying blue blobs. However, no blue blobs were found through a different citizen science search covering the entirety of Fornax (Mazziotti et al. in prep.). This may be because the Fornax cluster is dynamically older than Virgo or because it is lower mass. Any other cluster would be too distant to achieve the same level of deep coverage as NGVS with current telescopes.
Thus, for the immediate future, the Virgo cluster appears to be the only location where these objects can be discovered and studied. The completion of a visual search of such an extensive data set, which otherwise would have been a long and impractical process for a research team, highlights the significant impact and effectiveness of citizen science.

Confirmation of additional blue blobs from our candidate list will require further spectral follow-up targeting the \halpha and \hi emission lines. We are actively pursuing these with HET and GBT, respectively. To form a better understanding of when these objects begin forming stars and their subsequent evolution will require deep imaging of their resolved stellar populations, a task which is best-suited to JWST.

\begin{acknowledgments}
The science in this publication was made possible by the participation of over 1400 volunteers in the Blobs and Blurs citizen science project. We gratefully acknowledge their contribution to this work and list them all at \url{https://www.zooniverse.org/projects/mike-dot-jones-dot-astro/blobs-and-blurs-extreme-galaxies-in-clusters/about/results}.
This publication uses data generated via the Zooniverse.org platform, development of which is funded by generous support, including from the National Science Foundation, NASA, the Institute of Museum and Library Services, UKRI, a Global Impact Award from Google, and the Alfred P. Sloan Foundation.
We thank the NGVS and GALEX teams who produced the public legacy data sets that made this work possible.
This paper is based on observations obtained with MegaPrime/MegaCam, a joint project of CFHT and CEA/IRFU, at the Canada–France–Hawaii Telescope (CFHT), which is operated by the National Research Council (NRC) of Canada, the Institut National des Sciences de l'Univers of the Centre National de la Recherche Scientifique (CNRS) of France, and the University of Hawaii.
This research used the facilities of the Canadian Astronomy Data Centre operated by the National Research Council of Canada with the support of the Canadian Space Agency. 
This work used images from the Dark Energy Camera Legacy Survey (DECaLS; Proposal ID 2014B-0404; PIs: David Schlegel and Arjun Dey). Full acknowledgment at \url{https://www.legacysurvey.org/acknowledgment/}.
This work has used the NASA/IPAC Extragalactic Database (NED), which is funded by the National Aeronautics and Space Administration and operated by the California Institute of Technology.
This research has made use of the NASA/IPAC Infrared Science Archive, which is funded by the National Aeronautics and Space Administration and operated by the California Institute of Technology.
DJS and the Arizona team acknowledges support from NSF grant AST-2205863.

\end{acknowledgments}

%

\vspace{5mm}
\facilities{CFHT, GALEX, Blanco, HET, Arecibo, ROSAT, HST, IRSA, IRAS}


\software{\href{https://www.astropy.org/index.html}  
               {\texttt{Astropy}} \citep{astropy2013,astropy2018}, \href{https://reproject.readthedocs.io/en/stable/}{\texttt{reproject}} \citep{reproject}, \href{https://sites.google.com/cfa.harvard.edu/saoimageds9}{\texttt{DS9}} \citep{DS9}, \href{https://cartavis.org/}{\texttt{CARTA}} \citep{CARTA}, \href{https://www.aperturephotometry.org/}{\texttt{Aperture Photometry Tool}}, \href{https://matplotlib.org/}{\texttt{matplotlib}} \citep{matplotlib}, \href{https://numpy.org/}{\texttt{numpy}} \citep{numpy}, \href{https://scipy.org/}{\texttt{scipy}} \citep{scipy1,scipy2},  \href{https://pandas.pydata.org/}{\texttt{pandas}} \citep{pandas1,pandas2}, \href{https://github.com/grzeimann/Panacea}{\texttt{Panacea}},\href{https://github.com/grzeimann/LRS2Multi}{\texttt{LRS2Multi}}, \href{https://dust-extinction.readthedocs.io}{\texttt{dust\_extinction}}.}


\appendix

\section{Jellyfish structures Identified}\label{jellystablesec}
All the jellyfish structures identified through the citizen science search and their likely associated galaxies are listed in Table \ref{jellytable}. A total of 56 jellyfish structures associated with 44 independent galaxies were identified. Each jellyfish structure was visually checked in the Legacy Viewer to see if it was near the parent galaxy.

\begin{table}[!ht]
\caption{All Jellyfish candidates found in the main Virgo cluster (\S\ref{candidate classification}) through the citizen science search. The columns are as follows: (1) Assigned name; (2) AGC name (if present); (3) Other names in a major catalog (if present); (4) (5) spatial coordinates in J2000.}
\resizebox{0.47\textwidth}{!}{%
\begin{tabular}{c c c c c}
\toprule
Name        & AGC    & other name   & RA       & DEC      \\ \hline
BC1213+1360 & 007216 & IC 3044      & 12:12:48 & +13:59:48 \\
BC1213+1256 & 007220 & IC 3046a       & 12:13:04 & +12:56:14 \\
BC1214+1453 & 007231 & NGC 4192     & 12:13:44 & +14:52:42 \\
BC1218+1223 & 007326 & IC 3105      & 12:17:31 & +12:22:57 \\
BC1219+1427 &        & NGC 4254     & 12:18:33 & +14:26:45 \\
BC1219+1422 &        & NGC 4254     & 12:18:38 & +14:21:45 \\
BC1219+1429 &        & NGC 4254     & 12:18:48 & +14:28:48 \\
BC1219+1423 &        & NGC 4254     & 12:19:09 & +14:22:55 \\
BC1219+1253 & 220351 & VCC 328      & 12:19:14 & +12:53:19 \\
BC1219+1452 & 007365 & NGC 4262        & 12:19:26 & +14:51:41 \\
BC1222+1439 & 007418 & NGC 4302        & 12:21:40 & +14:38:40 \\
BC1223+1542 &        & NGC 4321        & 12:23:02 & +15:41:59 \\
BC1223+1344 & 220478 & VCC 618      & 12:23:08 & +13:44:04 \\
BC1223+1121 & 007456 & NGC 4330        & 12:23:11 & +11:20:42 \\
BC1223+1121 & 007456 & NGC 4330        & 12:23:15 & +11:21:19 \\
BC1225+1630 & 007507 & NGC 4383        & 12:25:24 & +16:29:59 \\
BC1225+1626 & 007507 & NGC 4383        & 12:25:29 & +16:25:52 \\
BC1226+0715 & 007513 & IC 3322a       & 12:25:40 & +07:14:50  \\
BC1226+1241 & 007520 & NGC 4388        & 12:25:48 & +12:40:34 \\
BC1227+1555 & 007563 & IC 3365        & 12:27:19 & +15:55:06 \\
BC1227+1303 & 007574 & NGC 4438        & 12:27:25 & +13:02:40 \\
BC1228+1303 & 007574 & NGC 4438        & 12:27:35 & +13:02:35 \\
BC1228+0852 & 225023 & LEDA 1355351 & 12:27:48 & +08:52:28  \\
BC1228+1304 & 007574 & NGC 4438        & 12:27:52 & +13:03:45 \\
BC1228+0924 & 007587 & NGC 4445        & 12:28:20 & +09:24:01  \\
BC1229+0609 & 222309 & 042-123a     & 12:28:49 & +06:09:23  \\
BC1229+0960 & 220661 & IC 3412        & 12:29:23 & +09:59:48  \\
BC1230+0750 & 007627 & NGC 4470        & 12:29:39 & +07:50:24  \\
BC1230+1123 &        & IC3418       & 12:29:50 & +11:22:46 \\
BC1230+1036 & 007633 & IC 3425        & 12:29:53 & +10:36:11 \\
BC1230+1122 &        & IC 3418       & 12:29:57 & +11:22:16 \\
BC1230+0757 &        & UGC 7636     & 12:29:58 & +07:56:50  \\
BC1232+1404 & 007695 & IC 3476        & 12:32:24 & +14:03:35 \\
BC1232+1147 & 007686 & IC 3467        & 12:32:27 & +11:47:30 \\
BC1233+1403 & 007695 & IC 3476        & 12:32:32 & +14:02:33 \\
BC1233+1123 &        & IC 3481A     & 12:33:05 & +11:22:47 \\
BC1233+0842 &        & NGC 4519     & 12:33:25 & +08:42:29  \\
BC1234+0911 & 007711 & NGC 4522        & 12:33:39 &+09:10:48  \\
BC1234+0807 &        & NGC 4535     & 12:34:09 & +08:07:08  \\
BC1234+0627 & 007726 & NGC 4532        & 12:34:21 & +06:27:07  \\
BC1236+1302 &        & NGC 4569        & 12:36:24 & +13:02:24 \\
BC1236+1313 &        & NGC 4569        & 12:36:27 & +13:12:36 \\
BC1237+1112 & 007777 & NGC 4567        & 12:36:31 & +11:12:19 \\
BC1237+1307 &        & NGC 4569        & 12:36:33 & +13:06:38 \\
BC1237+1315 & 007784 & IC 3583        & 12:36:45 & +13:14:57 \\
BC1238+0834 & 220849 & 070-196      & 12:37:44 & +08:33:40  \\
BC1238+0705 & 007795 & 042-184      & 12:37:50 & +07:04:59  \\
BC1238+1029 & 007808 & IC 3608        & 12:38:29 & +10:28:39 \\
BC1239+1028 & 007808 & IC 3608        & 12:38:43 & +10:28:23 \\
BC1239+0758 & 007822 & IC 3617        & 12:39:21 & +07:58:08  \\
BC1243+1059 & 226144 &              & 12:42:36 & +10:59:05 \\
BC1243+1622 &        & NGC 4651       & 12:43:26 & +16:22:27 \\
BC1244+1307 & 007902 & NGC 4654        & 12:44:06 & +13:06:32 \\
BC1252+1206 & 008007 & NGC 4746        & 12:51:50 & +12:05:44 \\
BC1253+1424 & 221085 & GR21         & 12:53:00 & +14:24:28 \\
BC1253+1238 & 223247 & KK169        & 12:53:11 & +12:38:14 \\
\hline
\end{tabular}%
}
\label{jellytable}
\end{table}

\section{Newly confirmed blue blobs}
\label{sec:HETspectra} 

\begin{figure*}
    \centering
    \begin{minipage}[t]{0.48\linewidth}
        \includegraphics[width=\linewidth]{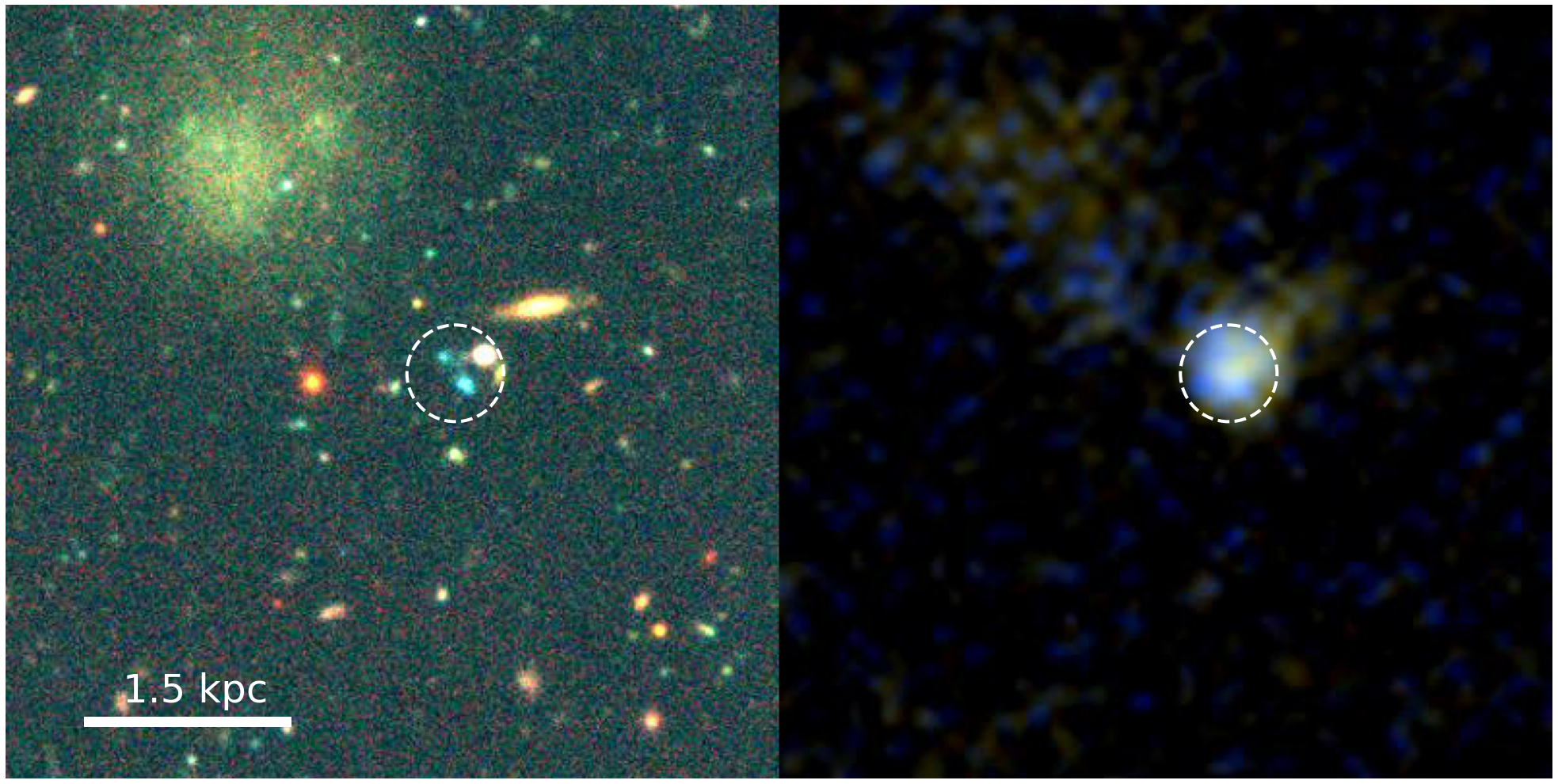}
        \centering (a) BC1224+1148 (BC7)
    \end{minipage}
    \hspace{0.01\linewidth}
    \begin{minipage}[t]{0.48\linewidth}
        \includegraphics[width=\linewidth]{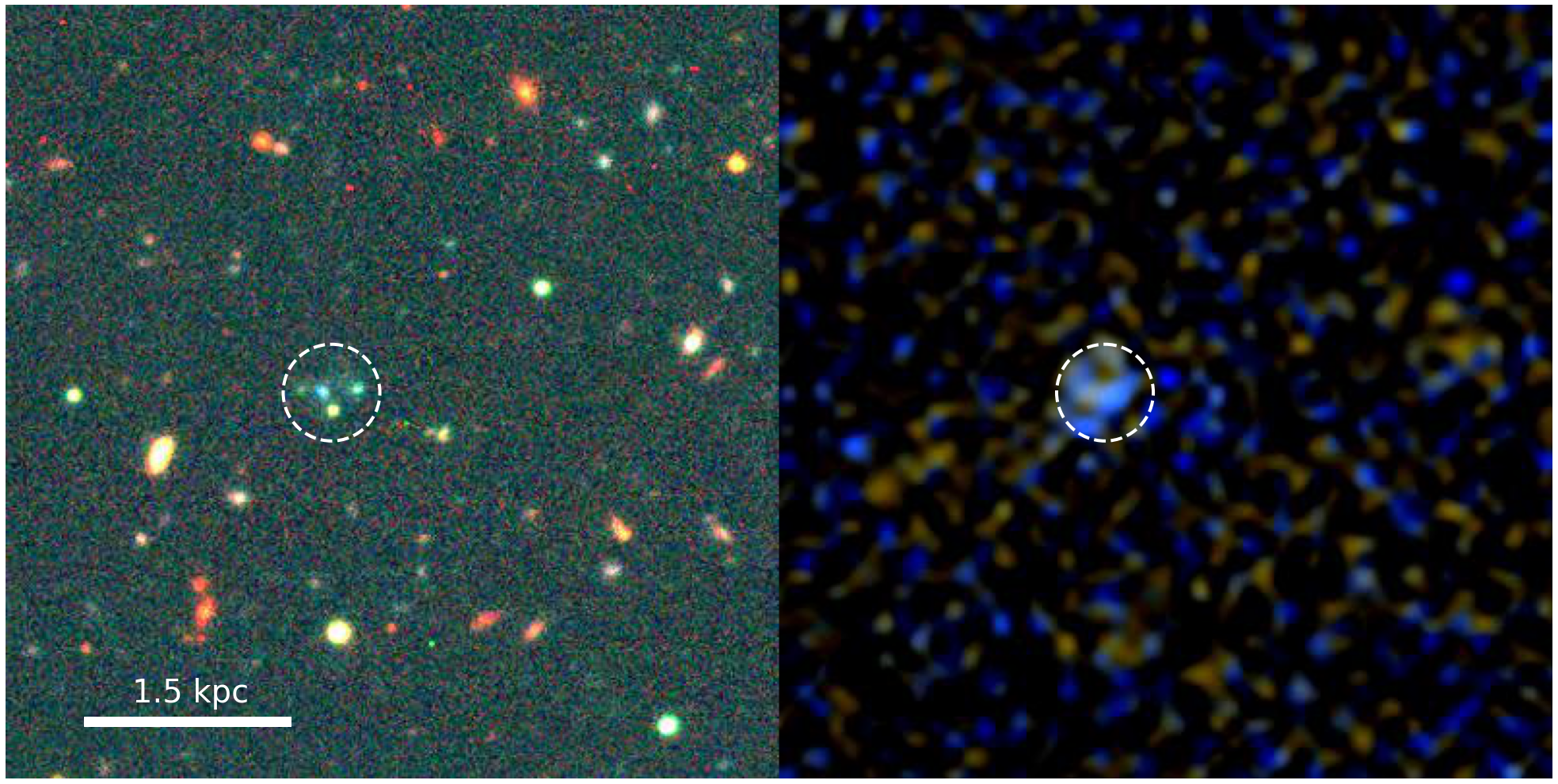}
        \centering (b) BC1247+1021 (BC12)
    \end{minipage}
    
    \vspace{1em} 
    
    \begin{minipage}[t]{0.48\linewidth}
        \includegraphics[width=\linewidth]{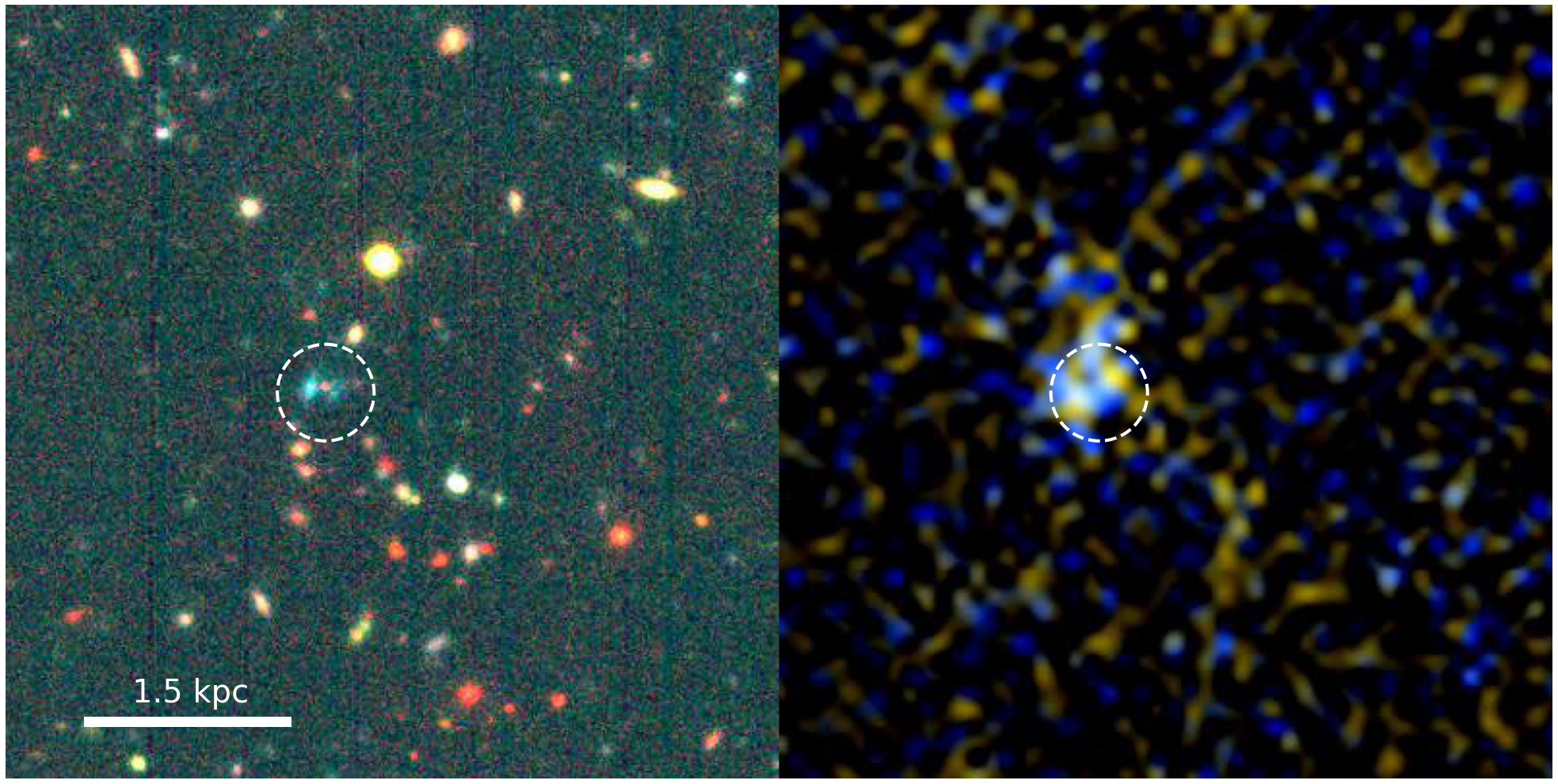}
        \centering (c) BC1242+0841 (BC13)
    \end{minipage}
    \hspace{0.01\linewidth}
    \begin{minipage}[t]{0.48\linewidth}
        \includegraphics[width=\linewidth]{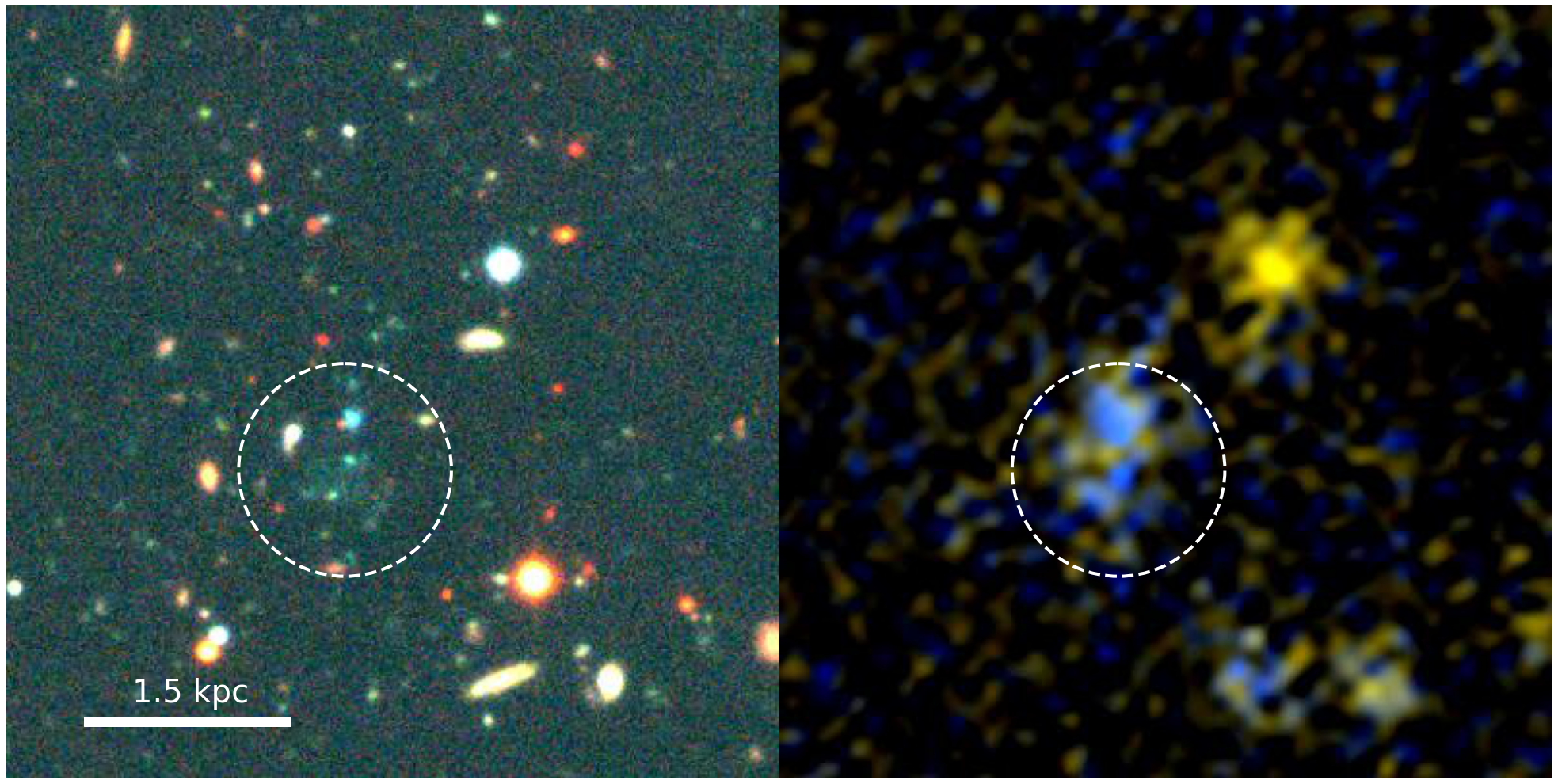}
        \centering (d) BC1236+0801 (BC17)
    \end{minipage}
    
    \vspace{1em}
    
    \begin{minipage}[t]{0.48\linewidth}
        \includegraphics[width=\linewidth]{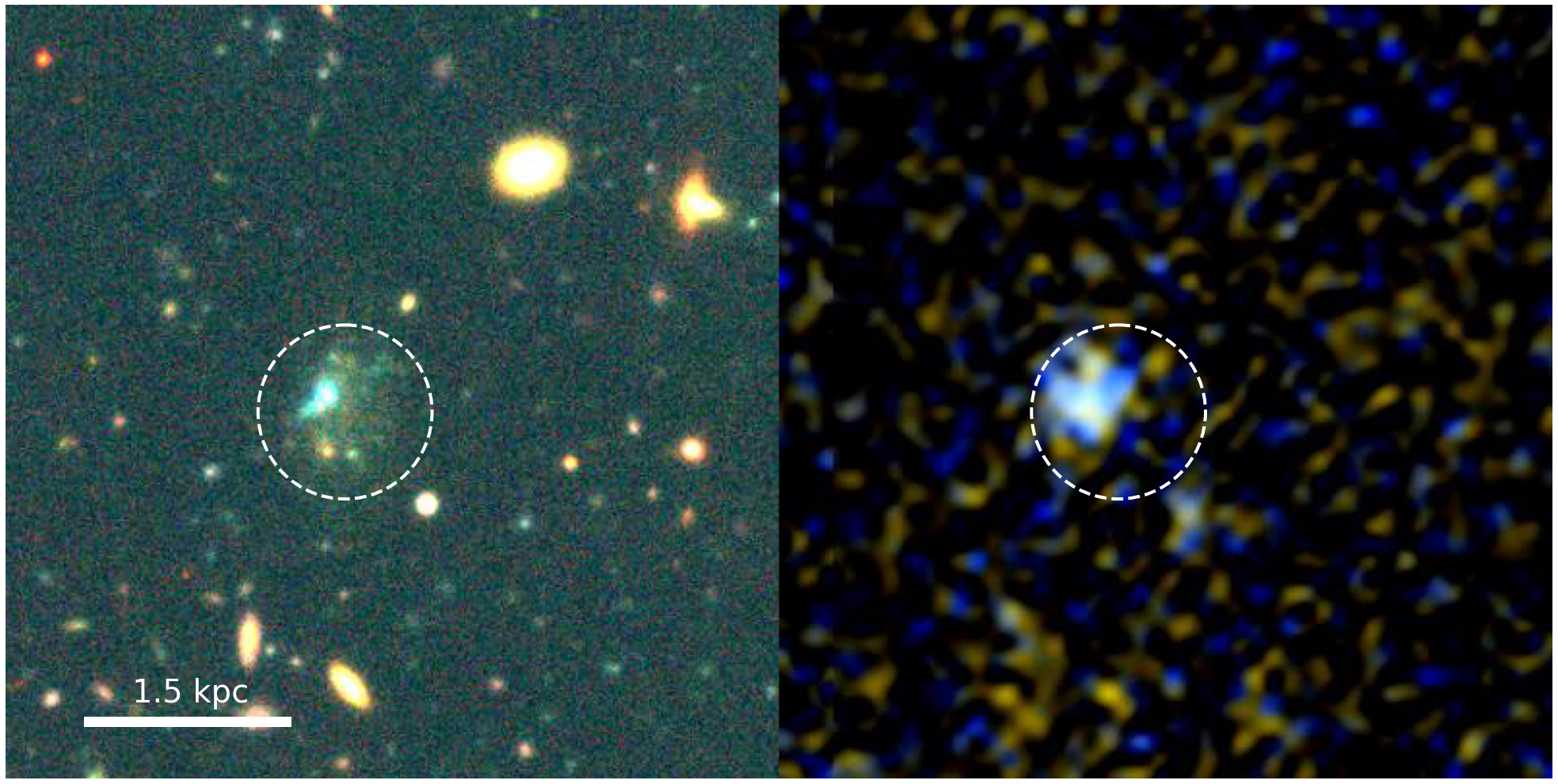}
        \centering (e) BC1230+0839 (BC25)
    \end{minipage}
    \hspace{0.01\linewidth}
    \begin{minipage}[t]{0.48\linewidth}
        \includegraphics[width=\linewidth]{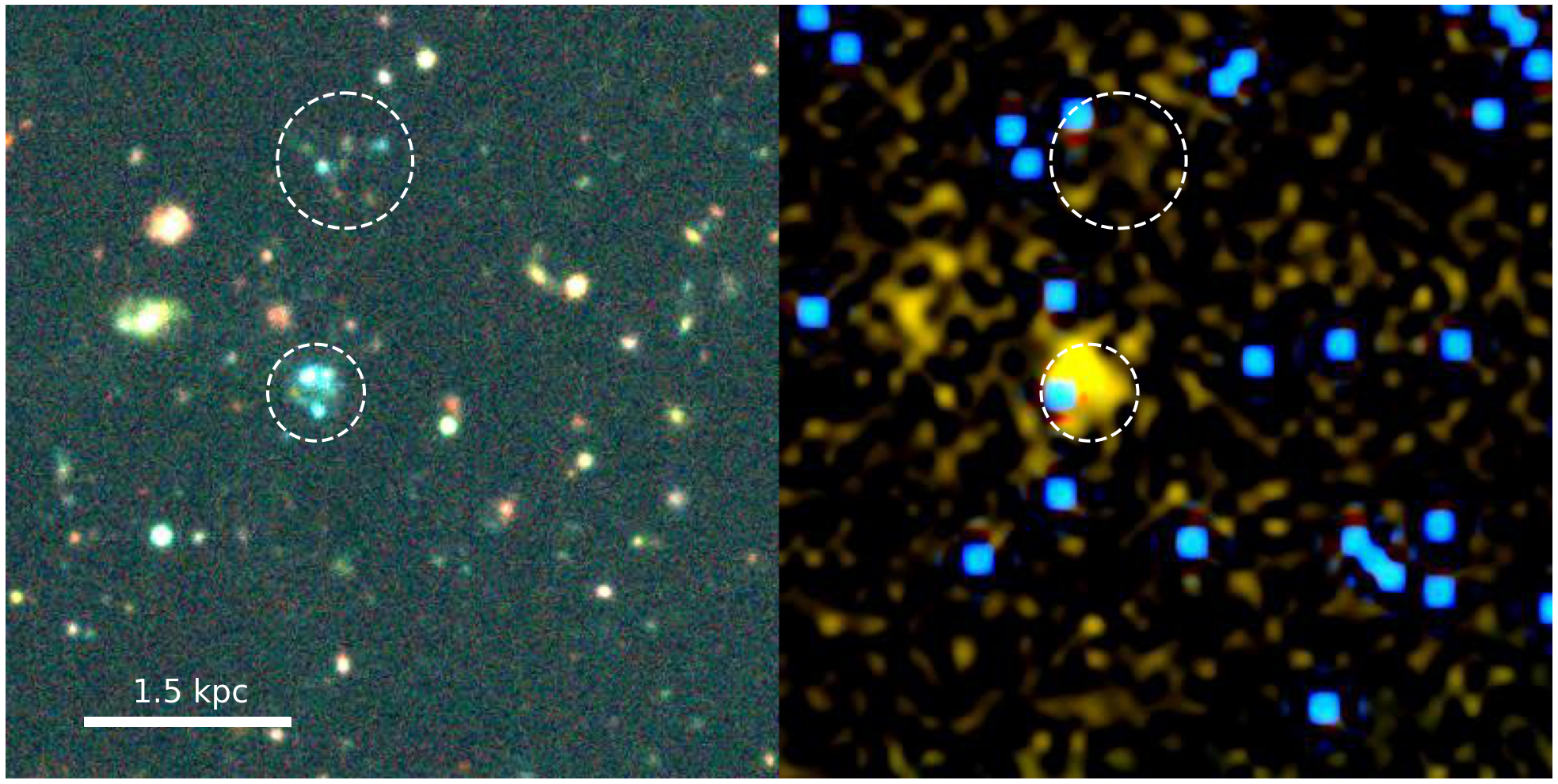}
        \centering (f) BC1226+1429 (BC30)
    \end{minipage}

    \caption{NGVS and GALEX cutout of each of the newly confirmed blue blobs. These blue blobs were confirmed to have membership with the Virgo cluster using emission line spectroscopy with HET.}
    \label{fig:ngvs}
\end{figure*}

\begin{figure*}
    \centering
    \includegraphics[width=0.49\linewidth]{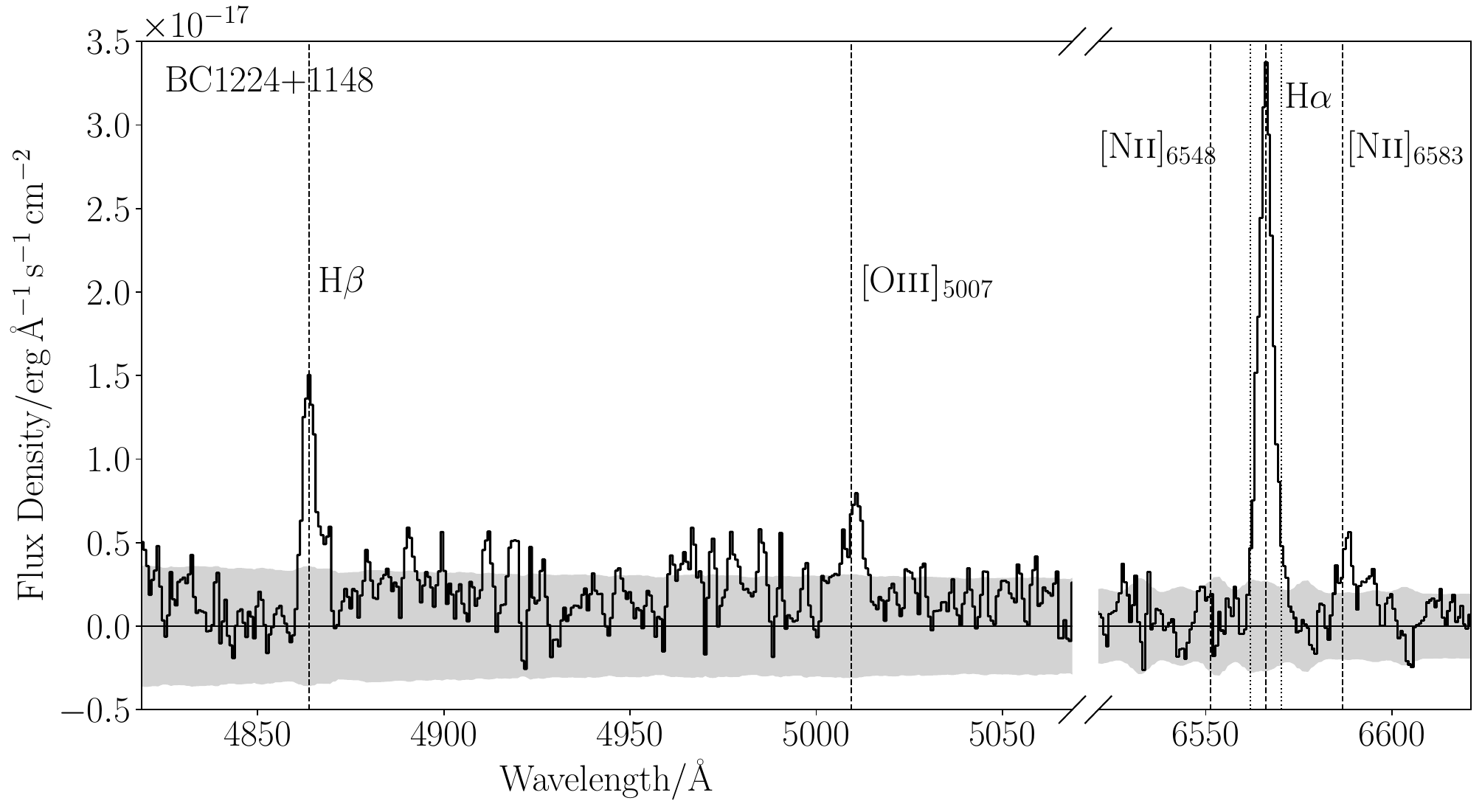}
    \includegraphics[width=0.49\linewidth]{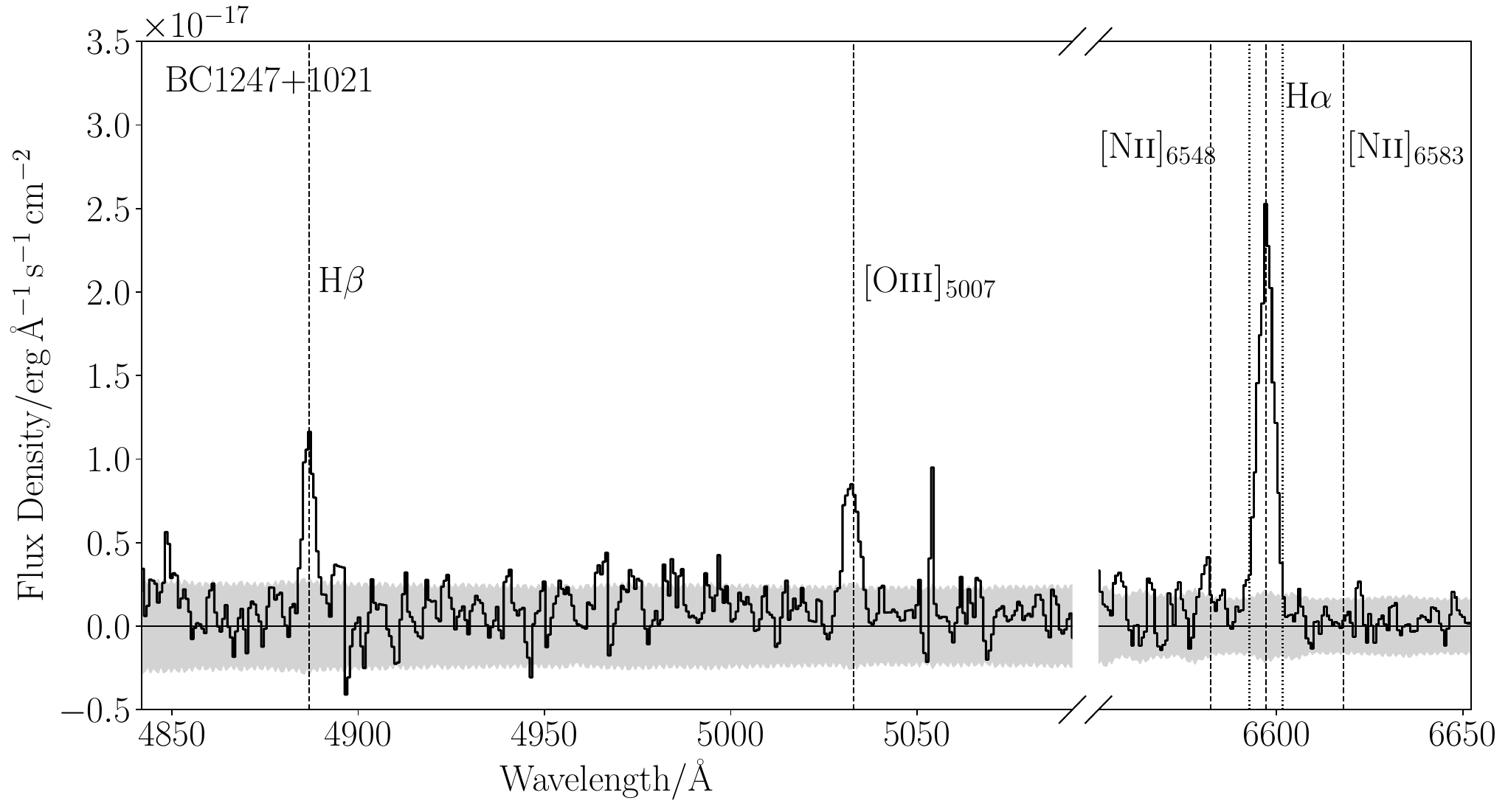}
    \includegraphics[width=0.49\linewidth]{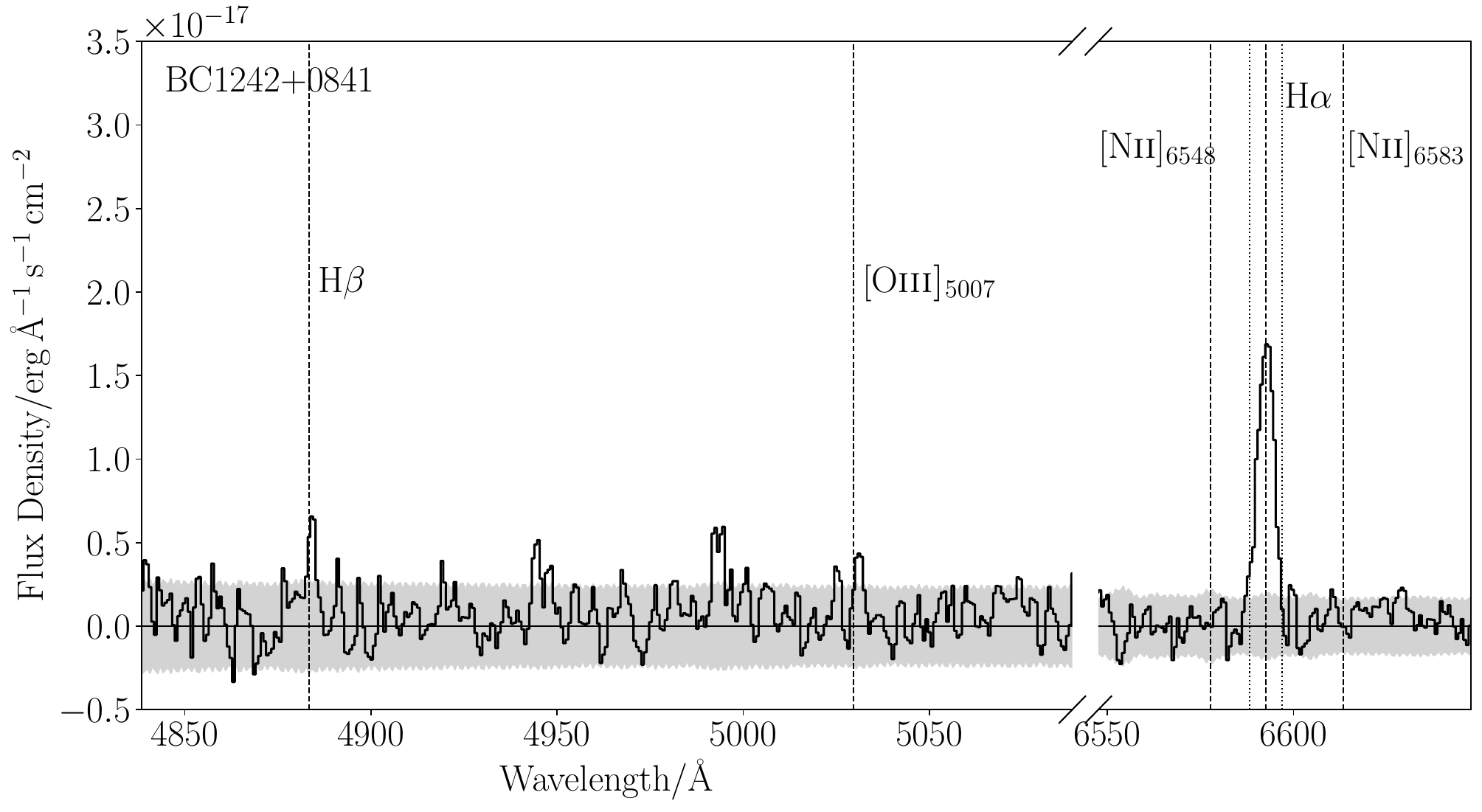}
    \includegraphics[width=0.49\linewidth]{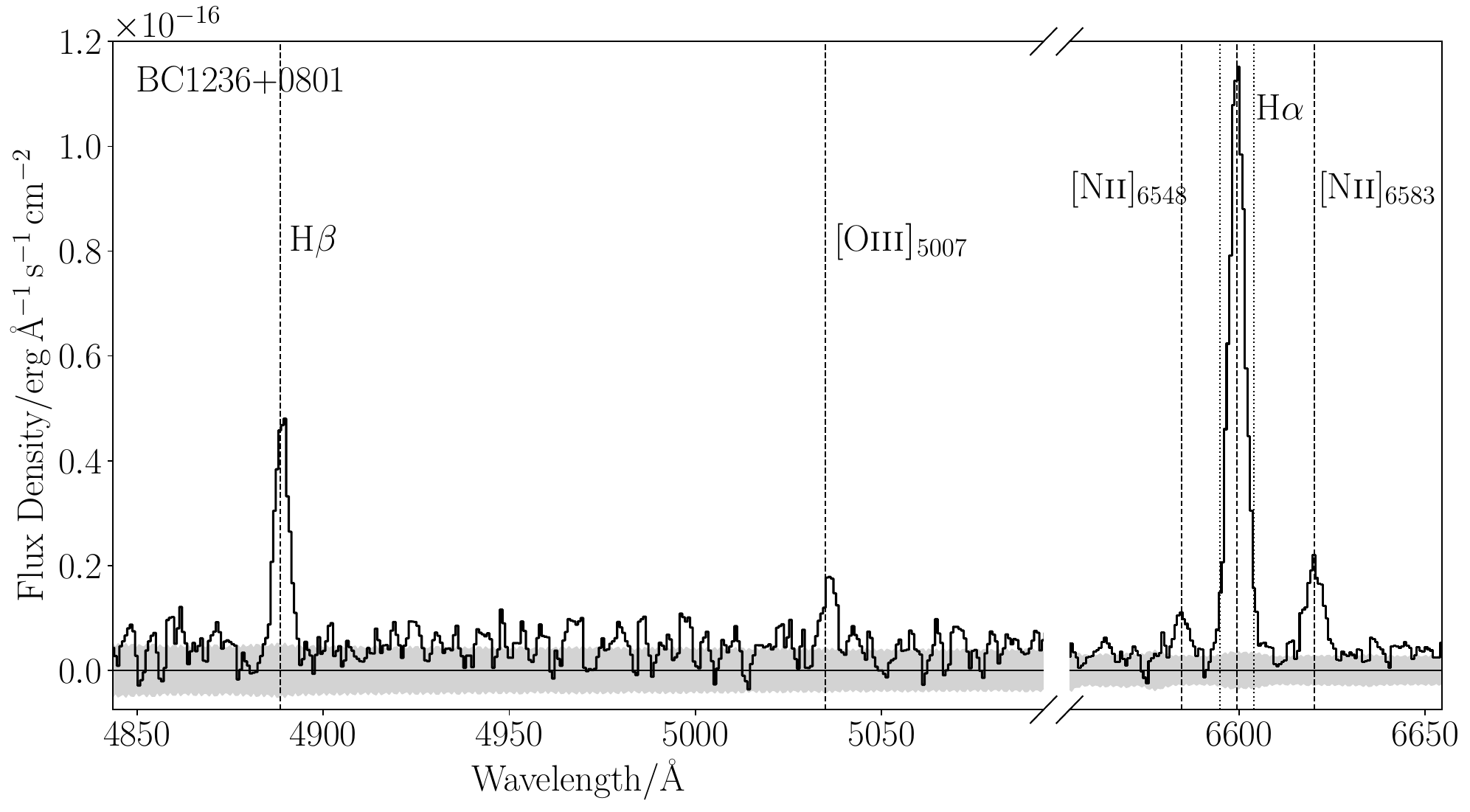}
    \includegraphics[width=0.49\linewidth]{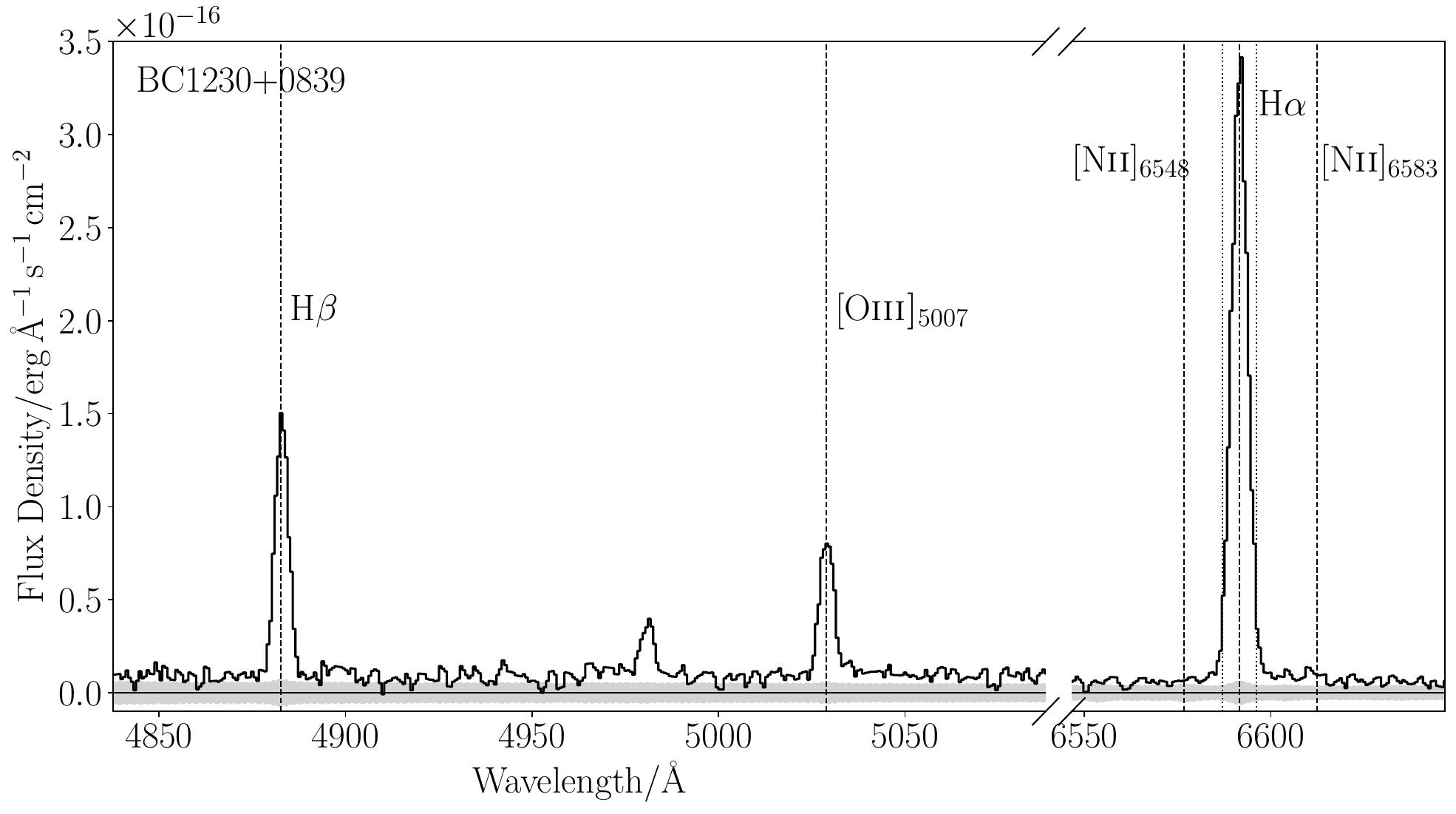}
    \includegraphics[width=0.49\linewidth]{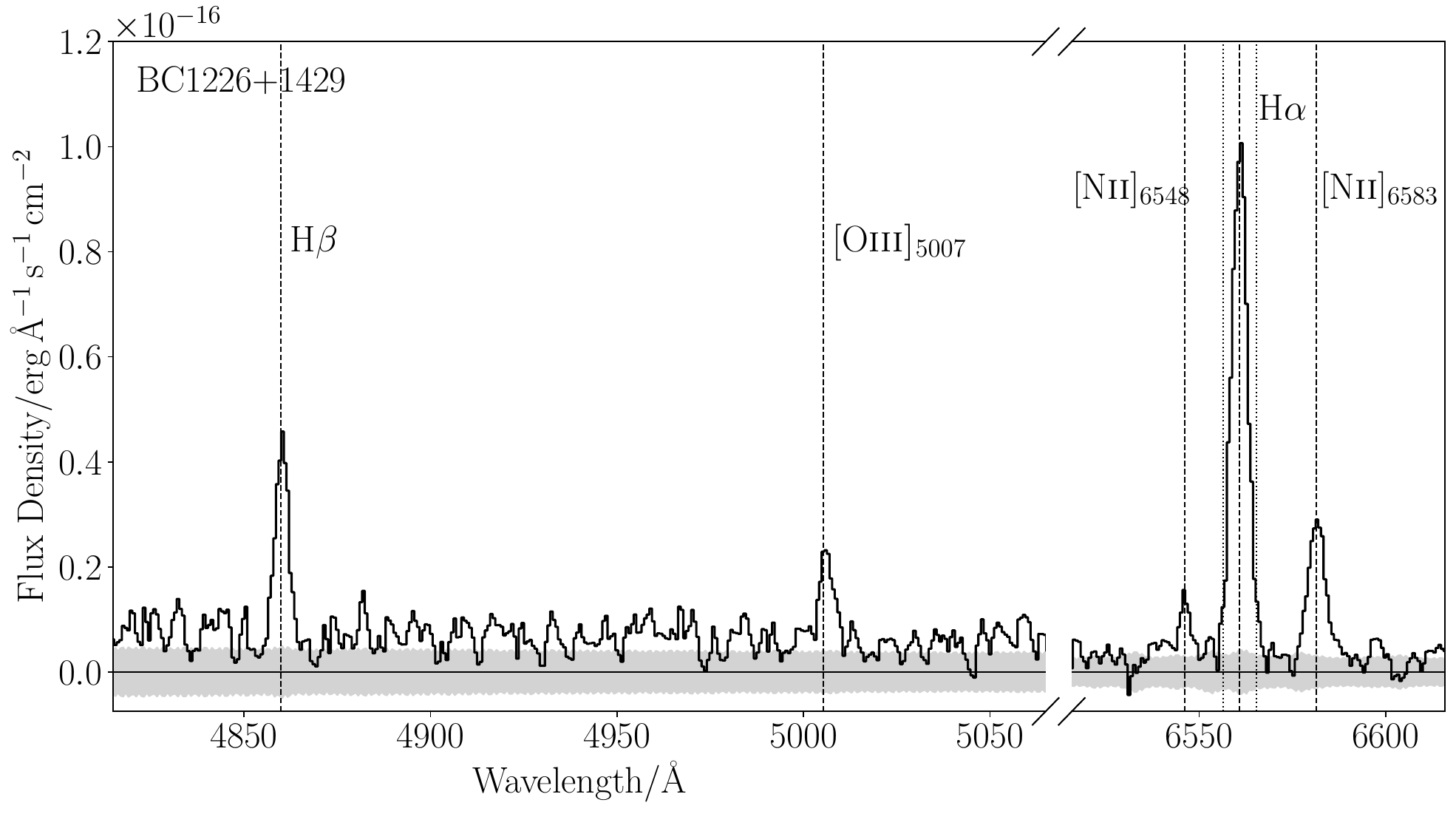}
    \caption{HET spectra of the newly identified blue blobs. The grey bands indicate the 1$\sigma$ uncertainties, and the vertical dashed lines show the observer frame wavelengths of various emission lines (based on the fitted redshift of H$\alpha$). The vertical dotted lines on either side of the H$\alpha$ line indicate the 1$\sigma$ width of the Gaussian fit to the line.}
    \label{fig:HETspectra}
\end{figure*}

Figure~\ref{fig:ngvs} shows the NGVS and GALEX cutouts of the six new velocity-confirmed blue blobs identified from our Zooniverse search (Table~\ref{table1}). In Figure~\ref{fig:HETspectra} we show the HET spectra used to confirm these blue blobs through radial velocity and metallicity measurements. They are displayed in the same order as in Figure~\ref{fig:ngvs}. In all cases H$\alpha$, H$\beta$, and [O{\sc iii}] lines are visible, and in most cases [N{\sc ii}] lines are also detected. 

\section{Phase space map of blue blobs}
\label{sec:phase_map}

\begin{figure}
    \centering
    \includegraphics[width=0.49\linewidth]{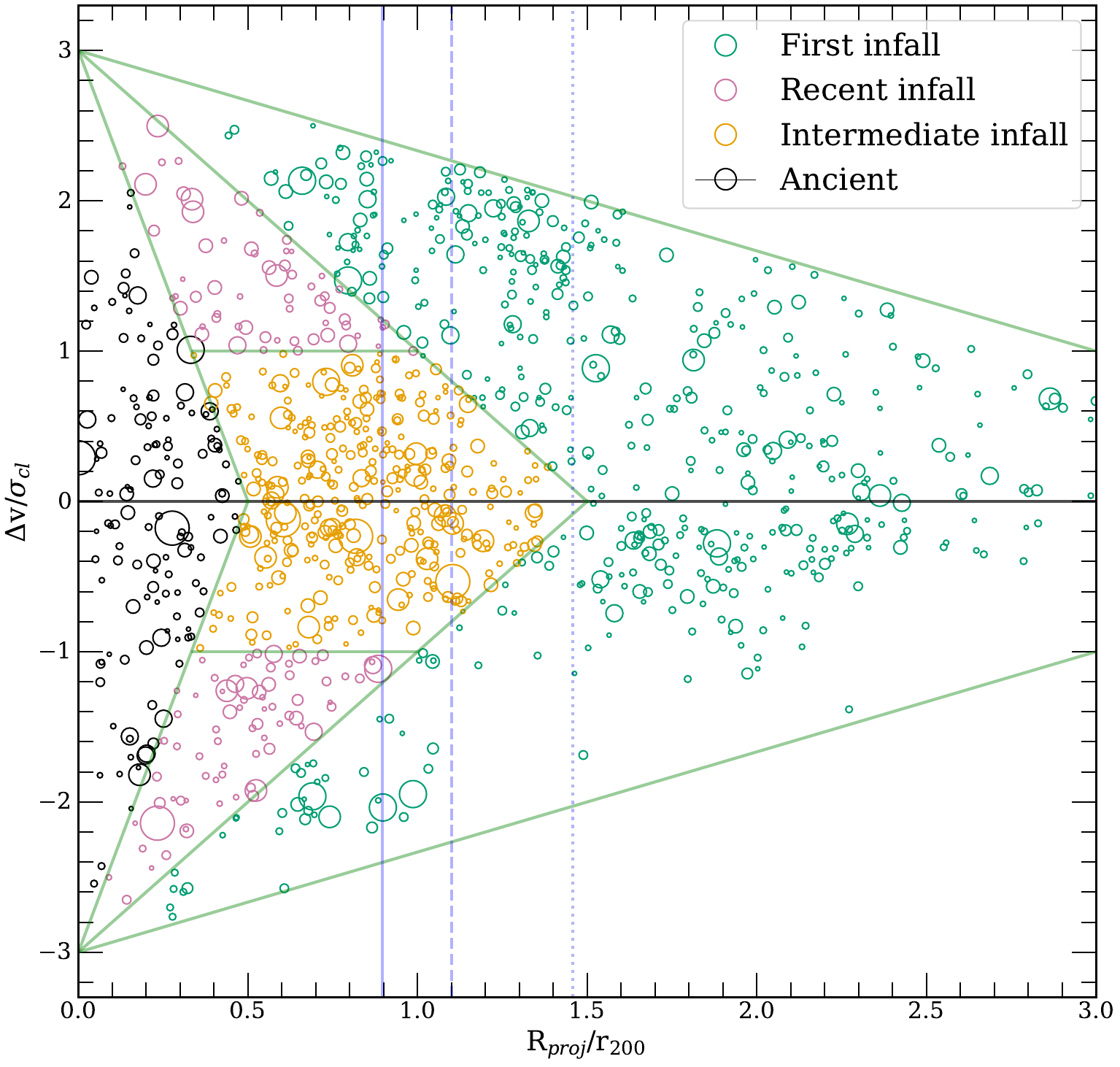}
    \includegraphics[width=0.49\linewidth]{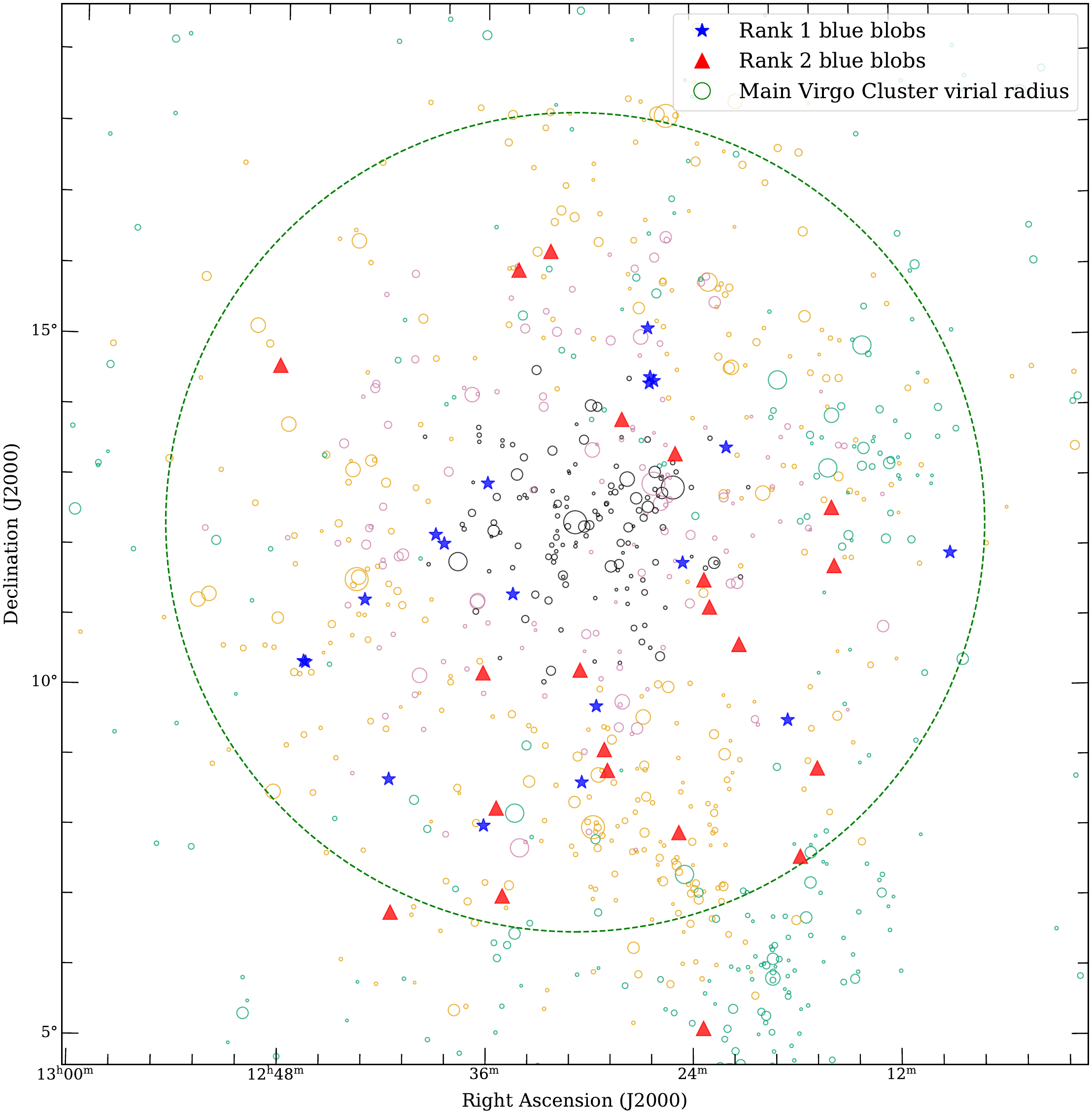}
    \caption{Left: Projected phase-space diagram of EVCC galaxies colored according to the stage of their orbit defined in \citet{mun} and shown in Fig. \ref{fig:RPS}. The area of each EVCC galaxy circle is determined by the total r-band flux of the represented galaxy. Right: We re-plot Figure \ref{fig:distBC} but now use the same color for EVCC galaxies used in the adjacent plot. Blue blobs are mostly co-spatial with EVCC galaxies in the recent or intermediate infall stage.}
    \label{fig:RPS_2d}
\end{figure}

The left panel of Figure \ref{fig:RPS_2d} is a modification of Figure \ref{fig:RPS}, where the EVCC galaxies are color-coded depending on their infall category. The right panel shows the spatial distribution of these galaxies, using the same color scheme as the adjacent plot, along with the position of BCs. To reiterate, we see that rank 1 BCs are mostly associated with galaxies in intermediate infall regions, hinting that their parent galaxies must have had multiple crossings across the cluster center. This is also true for the majority of jellyfish structures identified.

\bibliography{sample631}{}
\bibliographystyle{aasjournal}



\end{document}